\documentclass[11pt]{article}

\usepackage{amsfonts,epsfig,fullpage}

\usepackage[table]{xcolor}
\usepackage{multirow}
\usepackage{array}
\newcolumntype{?}{!{\vrule width 1pt}}

\newtheorem{theo}{Theorem}

\newtheorem{lem}{Lemma}
\newtheorem{claim}[lem]{Claim}
\newtheorem{coro}[lem]{Corollary}

\newtheorem{define}{Definition}

\newtheorem{alg}{Algorithm}
\newtheorem{proc}{Procedure}

\newcommand{\BE}{\begin{enumerate}} \newcommand{\EE}{\end{enumerate}}
\newcommand{\BI}{\begin{itemize}} \newcommand{\EI}{\end{itemize}}
\newcommand{\BDes}{\begin{description}}\newcommand{\EDes}{\end{description}
}
\newcommand{\BT}{\begin{theo}} \newcommand{\ET}{\end{theo}}
\newcommand{\BL}{\begin{lem}} \newcommand{\EL}{\end{lem}}
\newcommand{\BD}{\begin{define}} \newcommand{\ED}{\end{define}}
\newcommand{\BCM}{\begin{claim}} \newcommand{\ECM}{\end{claim}}
\newcommand{\BC}{\begin{coro}} \newcommand{\EC}{\end{coro}}
\newcommand{\BA}{\begin{alg}} \newcommand{\EA}{\end{alg}}
\newcommand{\BP}{\begin{proc}} \newcommand{\EP}{\end{proc}}

\def\FullBox{\hbox{\vrule width 8pt height 8pt depth 0pt}}
\newcommand{\qed}{\;\;\;\FullBox}
\newenvironment{proof}{\noindent{\bf Proof:~~}}{\(\qed\)}
\newcommand{\BPF}{\begin{proof}} \newcommand {\EPF}{\end{proof}}
\newenvironment{proofof}[1]{\noindent{\bf Proof of {#1}:~~}}{\(\qed\)}
\newcommand{\BPFOF}{\begin{proofof}} \newcommand {\EPFOF}{\end{proofof}}

\newcommand{\BEQ}{\begin{equation}} \newcommand{\EEQ}{\end{equation}}
\newcommand{\BEQN}{\begin{eqnarray}}\newcommand{\EEQN}{\end{eqnarray}}

\newlength{\saveparindent}
\newlength{\saveparskip}
\newtheorem{exmp}{Example}

\begin{document}

\title{The Augmentation Property of Binary Matrices \\for the Binary and Boolean Rank }
\author{
    Michal Parnas \\
    The Academic College\\
    of Tel-Aviv-Yaffo \\
    Tel-Aviv, {\sc Israel} \\
    {\tt michalp@mta.ac.il}
\and
    Adi Shraibman \\
   The Academic College\\
    of Tel-Aviv-Yaffo \\
    Tel-Aviv, {\sc Israel} \\
    {\tt adish@mta.ac.il}
}

\maketitle

\begin{abstract}
We define the Augmentation property for binary matrices with respect to different rank functions.
A matrix $A$ has the Augmentation property for a given rank function, if
for any subset of column vectors $x_1,...,x_t$ for for which the rank of $A$ does not increase when augmented
separately with each of the vectors $x_i$, $1\leq i \leq t$, it also
holds that the rank does not increase when augmenting $A$ with all vectors $x_1,...,x_t$ simultaneously.
This property holds trivially for the usual linear rank over the reals,
but as we show, things change significantly when considering the binary and boolean rank of a matrix.

We prove a necessary and sufficient condition for this property to hold under the binary and boolean rank of binary matrices.
Namely, a matrix has the Augmentation property for these rank functions
if and only if it has a unique base that spans all other bases of the matrix
with respect to the given rank function.
For the binary rank, we also present a concrete characterization of a family of matrices
that has the Augmentation property. This characterization is based on the possible types
of linear dependencies between rows of $V$, in optimal binary decompositions of the matrix as $A=U\cdot V$.

Furthermore, we use the Augmentation property to construct
simple families of matrices, for which there is a gap between their real and binary rank
and between their real and boolean rank.
\end{abstract}

\section{Introduction}
The notion of the rank of a matrix over $\mathbb{R}$, or the rank over any field for that matter,
is well understood. Many powerful techniques, most of which come from linear algebra,
were devised over the years, which enable us to prove strong results involving the
real rank of a matrix.
For this reason the rank appears in numerous situations in different areas of mathematics,
including the field of theoretical computer science.
In Section~\ref{sec:binary_and_boolean_rank_communication_complexity}
we illustrate this by elaborating on the role of the rank function in the field of communication complexity,
a role which started with a lower bound of Melhorn and Schmidt~\cite{mehlhorn1982vegas}.
As in communication complexity, the linear algebraic tools and the many properties of the rank function,
are those that make it attractive to use in many applications.

Denote by $R_{\mathbb{R}}(A)$ the rank of a real $n \times m$ matrix $A$ over $\mathbb{R}$.
Naming just a few of the useful properties of $R_{\mathbb{R}}(A)$, we have:

\begin{itemize}
\item $R_{\mathbb{R}}(A)$ is equal to the minimal size of a spanning set among the rows/columns of $A$.

\item $R_{\mathbb{R}}(A)$ is equal to the maximal size of an independent set among the rows/columns of $A$.

\item $R_{\mathbb{R}}(A)$ is equal to the minimal $k$ for which there is an $n \times k$ matrix $U$ and a $k\times m$
matrix $V$ such that $A=U\cdot V$.

\item Sub-additivity: $R_{\mathbb{R}}(A+B) \le R_{\mathbb{R}}(A) + R_{\mathbb{R}}(B)$.

\item Multiplicity under tensor product: $R_{\mathbb{R}}(A \otimes B) =
R_{\mathbb{R}}(A)\cdot R_{\mathbb{R}}(B)$.

\end{itemize}

Other rank functions, whose behavior is far less understood, are also defined and studied in the literature \cite{BGP,Beasley,Caen,Cohen,GregoryPullman,Watson,Zhong}.
The usual definition is as follows.
Let $(S,+_S,\cdot_S)$ be a triplet, where $S$ is a basic set of elements, and $+_S$ and $\cdot_S$ are two
basic operations on pairs of elements from $S$.
In many cases $(S,+_S,\cdot_S)$ is a semi-ring (see e.g. Gregory and Pullman~\cite{GregoryPullman}).

Then given an $n \times m$ matrix $A$, the rank function $R_S(A)$ is defined as the minimal $k$ such that there is a
decomposition $A = U\cdot V$, where $U$ is an $n \times k$ matrix, $V$ is a $k \times m$ matrix
and the entries of $U,V$ are from $S$.
The matrix product $U\cdot V$ is defined as usual,
using $+_S,\cdot_S$ as addition and multiplication between elements of the matrices.
The product $U\cdot V$ is called an {\em Optimal Decomposition} of $A$ for the given rank function $R_S(A)$,
and $k$ is the {\em size} of the decomposition.

With a slight abuse of notation, when the context is clear, we will simply write $+,\cdot$ instead of $+_S$ and $\cdot_S$.
Thus for example the usual rank $R_{\mathbb{R}}(A)$ is defined over $(\mathbb{R},+,\cdot)$.

When $(S, +_S,\cdot_S)$ is a field, or at least a ring, $R_S(A)$ has a strong
structure and many tools for estimating the rank are provided by linear algebra. When
$(S, +_S,\cdot_S)$ is a semi-ring, or has even less structure, most of these tools break down.

Some examples of rank functions $R_S(A)$, where $(S, +_S,\cdot_S)$ is a semi-ring,
that are used in theoretical computer science are:
\begin{itemize}
\item The {\em non-negative rank} which is defined over $(\mathbb{R}^+,+,\cdot)$,
where $+,\cdot$ are the regular addition and multiplication in $\mathbb{R}$.
\item The {\em Binary rank} which is defined over $(\mathbb{B},+,\cdot)$,
where $\mathbb{B}=\{0,1\}$, and again $+,\cdot$ are the standard addition and multiplication.
Note that here the basic set
is not closed under addition.
\item
The {\em Boolean rank}, defined over $(\mathbb{B},+_{\mathbb{B}},\cdot_{\mathbb{B}})$.
The operations are the boolean operations, i.e.
$0+_{\mathbb{B}}x = x+_{\mathbb{B}}0 = x$, $1+_{\mathbb{B}}1=1$, $ 1\cdot_{\mathbb{B}} 1 = 1$ and
$x \cdot_{\mathbb{B}} 0 = 0 \cdot_{\mathbb{B}} x = 0$.
\end{itemize}
Note that the basic set for both the binary and boolean rank  is $\mathbb{B}=\{0,1\}$,
and the only difference is in the definition of the addition operation,
where in the binary case $1+1=2$ and the sum is outside the basic set, whereas in the boolean case $1+ 1 = 1$.
Therefore, in order to distinguish between the two, we will denote by $binary$ the basic set in the binary setting,
and by $bool$ the basic set in the boolean setting.
Thus when discussing the binary rank it will always be with respect to $(binary,+,\cdot)$,
and when discussing the boolean rank it will be with respect to $(bool,+,\cdot)$,
where of course the basic operations are those defined above for each of these rank functions.
We will also denote an optimal decomposition in the binary setting as an {\em Optimal Binary Decomposition}
and in the boolean setting as an {\em Optimal Boolean Decomposition}.

\subsection{The Binary and Boolean Rank Functions in Communication Complexity}
\label{sec:binary_and_boolean_rank_communication_complexity}
The above rank functions arise in many mathematical scenarios and in different areas of
research.
One such scenario is that of communication complexity, described next.
For other areas of research that relate to these rank functions,
such as the Bi-clique edge partition/cover problems, Clustering and Tiling databases,
see for example~\cite{Gregory, Monson, Bein, geerts2004tiling}.

Communication complexity can be described as a game between two players, Alice and Bob.
Both Alice and Bob are familiar with some matrix $A=(a_{ij})$, usually a binary matrix,
and decide on some communication protocol for it.
Now Alice receives a row index $i$ and Bob receives a column index $j$, and their aim is to compute $a_{ij}$ by sharing as little
information as possible. The information is shared by transmitting bits ($0$ or $1$) in rounds, until both players know $a_{ij}$.
The {\em cost} of a protocol is the number of bits transmitted for the worst pair $(i,j)$ of row and column indices.
The {\em communication complexity} of $A$, denoted by $D(A)$,
is the minimal cost of a protocol for $A$. The non-deterministic communication complexity
of $A$, denoted by $N(A)$, is defined similarly with the exception that the communication is non-deterministic.
We omit the formal definition of non-deterministic communication complexity, as this is not needed throughout
the paper. We refer the interested reader to~\cite{KN97} for additional background on communication complexity.

A central notion in the study of communication complexity is that of a
{\em monochromatic combinatorial rectangle} in a binary matrix $A$.
A combinatorial rectangle in $A$ is essentially a sub-matrix of $A$, and a combinatorial rectangle
is called monochromatic if in this sub-matrix all entries have the same value.
Efficient communication protocols are strongly related to partitions of the entries of the matrix
into monochromatic combinatorial rectangles, as follows:

Let $Partition_1(A)$ be the size of a minimal partition of the $1$'s of $A$ (that is,
the subset of entries in $A$ that are equal to $1$), into monochromatic rectangles.
In a similar way let $Cover_1(A)$ be the size of a minimal cover of the $1$'s of $A$,
by monochromatic rectangles that may overlap.
Known results in communication complexity relate these notions to $D(A)$ and $N(A)$ as follows, for every binary matrix $A$
(see for example, Kushilevitz and Nisan~\cite{KN97} or Razborov~\cite{Razborov}):
\begin{itemize}
\item
$log_2 (Partition_1(A)) \leq D(A)  \leq O(log^2 Partition_1(A))$.
\item
$N(A) = log_2 (Cover_1(A)) $.
\end{itemize}
G\"{o}\"{o}s et al.~\cite{Goos} showed recently that there exists a matrix $A$ such that
$D(A) \geq \tilde{\Omega}(log^2 Partition_1(A))$.

Returning to our discussion about the various rank functions, and in particular the binary and boolean rank functions,
it holds that~\cite{Gregory}:
\begin{equation}
\label{binary boolean}
R_{binary}(A) = Partition_1(A), \ \ \ \ \ R_{bool}(A) = Cover_1(A).
\end{equation}
Thus, if we can compute $R_{binary}(A)$ for a given matrix $A$, we get a tight
estimate for $D(A)$, and computing $R_{bool}(A)$ determines exactly the non-deterministic communication complexity.
On the other hand, computing the real rank only gives a lower bound $D(A)\geq log_2 R_{\mathbb{R}}(A)$
(see~\cite{mehlhorn1982vegas}), whereas the current best upper bound that was proved by Lovett~\cite{Shachar}
is exponentially worse and is  $D(A) \leq \tilde{O}(\sqrt{R_{\mathbb{R}}(A)})$.
We note that the famous log-rank conjecture~\cite{Lovasz,NisanWigderson}
states that $D(A) \le \left(log R_{\mathbb{R}}(A)\right)^{O(1)}$,
and therefore of course the real rank may be a good estimate to $D(A)$, assuming the log-rank conjecture is true.


The log-rank conjecture could be proved in the future, in which case all the powerful tools
of linear algebra would apply directly to the communication complexity notions. Or it can
be disproved, and then the binary and boolean rank will remain as the only tight bound.
Since we do not know at the current time if the log rank is true or not, we offer
to pursue another line of research that is beneficial in both cases,
and that is to develop tools to deal with
the Boolean and Binary ranks and to investigate the properties of these rank functions.


\subsection{Our Results}
In this paper we examine the Binary and Boolean rank functions
(see also~\cite{Caen},\cite{GregoryPullman}),
and further restrict our attention to binary matrices $A$.
Note that the binary and boolean ranks are well defined for any binary matrix $A$,
since the trivial decomposition $A = A\cdot I_n$ is always possible,
where $n$ is the number of rows of $A$ and $I_n$ is the $n \times n$ identity matrix.

We focus on a specific linear algebraic property of  $R_{\mathbb{R}}()$,
and understanding how badly it breaks down for the binary and boolean rank.
Given a matrix $A$ and a column vector $x$, denote by $(A|x)$
the matrix that results by augmenting $A$ with $x$ as the last column.
Denote by $(A|x_1,...,x_t)$ the matrix that results from $A$ by augmenting it with the vectors
$x_1,...,x_t$.
The property we are concerned with, which we call the {\em Augmentation Property}, is:
\BD[Augmentation Property]
A matrix $A$ has the {\em Augmentation property} for a given rank function $R()$ if
for any subset of vectors $x_1,...,x_t$ for which $R(A|x_i) = R(A)$, for
$1\leq i \leq t$,
it also holds that $R(A|x_1,...,x_t)= R(A)$.
\ED

This property holds of course trivially with respect to the real rank, or the linear rank over any field.
This follows easily from linear algebra, as $R_{\mathbb{R}}(A|x_1,...,x_t) = R_{\mathbb{R}}(A)$
if and only if $x_i$ belongs to the subspace spanned by the columns of $A$ for every $i=1,\ldots,t$.
As we show, things change significantly when we use the binary or boolean rank.

Our main goal in this paper is therefore to examine what happens to the binary or boolean rank of a given binary matrix $A$,
after augmenting it with additional column vectors. We are interested in both positive and negative results.
Positive results, that is proving criteria under which the augmentation property holds for the binary or boolean rank,
provide tools to handle these rank functions.
Negative results help to clarify the limits of the linear algebraic tools,
and also sharpen the differences between these rank functions and the usual rank over $\mathbb{R}$.
Each such difference is a step towards understanding the truth in the log rank conjecture.

We first examine the Augmentation Property of a given binary matrix $A$ under the binary and boolean rank,
and prove a necessary and sufficient condition for the Augmentation property to hold.
The main technique used to prove this condition is a new concept of the {\bf Base Graph}
of a matrix that we define. The vertices of this graph are the different bases of the given matrix $A$
for a given rank function, and there is a directed edge from base $U$ to base $V$ if $U$ spans $V$.

The exact concepts of a base and spanning set will be defined later in the paper, but generally
speaking, a base is, as for the usual linear rank, a minimal subset of vectors that spans the columns
of the matrix, but here the vectors are binary, the coefficients in a linear combination are restricted to
$\{0,1\}$, and the $+,\cdot$ operations are the binary or the boolean operations (depending on the rank function in question).

Whereas this graph has a trivial structure for bases of the real rank,
since any base of the matrix $A$ spans all other bases of $A$ under the real rank,
it has interesting properties under the binary and boolean rank. Specifically it is always acyclic,
and this property is used to prove the following theorem:

\medskip\noindent {\bf Theorem 1:} {\em A binary matrix $A$ has the Augmentation property under the binary/boolean
rank if and only if  $A$ has a (unique) base that spans all other bases of $A$.}\\

Although Theorem 1 gives a sufficient and necessary condition for the Augmentation property,
it does not characterize the structure of the matrices
that have such a unique base that spans all other bases.
We thus study this question in search of such a characterization,
and achieve the following results for the binary rank.

We first observe that if $A$ is an $n\times m$ binary matrix, where $R_{binary}(A)= n$,
then $A$ has the augmentation property. The reason this holds is that the standard base
is a base of $A$ in this case. One can consider, more generally, bases that are
{\bf Disjoint in rows}.
That is, bases in which every two vectors in the base do not have a $1$ in the same row.
Indeed when the matrix does not have identical rows and has a base that is disjoint in rows,
then this base spans all other bases and so the Augmentation property holds for the binary rank. However,
as we will prove, if we omit the requirement that the matrix $A$ does not have identical rows,
then even the strong assumption of the existence of a disjoint in rows base is not sufficient to
guarantee the Augmentation property.

It is therefore interesting to find sufficient conditions for the augmentation property to hold,
even in the simple case that the matrix has a base that is disjoint in rows. To this end,
we define the {\bf Unique base rows sums} property for a matrix $A$, in which in any optimal binary decomposition
$X\cdot Y$ of $A$, the matrix $Y$ does not contain two subsets of rows whose sums are identical.
For matrices that posses this property we prove the following general result:

\medskip\noindent {\bf Theorem 2:} {\em
Let $A$ be a binary matrix with an optimal binary decomposition $A = X \cdot Y$,
such that the rows of $Y$ are rows of the matrix $A$. If, in addition, $A$
has the Unique base rows sums property, then $A$ has the Augmentation property for the binary rank.
}\\

Note that if a matrix $A$ has a base that is disjoint in rows, then $A$ has
an optimal binary decomposition $A = X \cdot Y$, such that the rows of $Y$ are rows of the matrix $A$.
This is because a matrix $A$ has a disjoint in rows base if and only if, after removing
rows that are all $0$ and identical rows, the resulting set of rows has a full rank. Theorem~2 holds more generally
for matrices such that after removing every row of $A$ that is linearly dependent on other rows
of $A$, the resulting set of rows has a full rank.

It is interesting to note that Theorem 2 does not hold for the boolean rank.

The concept of the base graph and the Augmentation property, allows us also to present a simple technique
to build a family of matrices in which there is a gap between the real rank and the binary rank and
between the real and boolean rank. Specifically we prove:

\medskip\noindent {\bf Theorem 3:} {\em
For every $d$, there exists a binary matrix $A$ such that $R_{binary}(A) = 4d$ and $R_{\mathbb{R}}(A) = 3d$.
}

\medskip\noindent {\bf Theorem 4:} {\em
For every $d$, there exists a binary matrix $A$ such that $R_{bool}(A) = 4d$ and $R_{\mathbb{R}}(A) = 3d$.
}\\

Although the gap achieved is not particularly large, we believe that the technique used to prove it is interesting
and hope that it can be used to prove stronger results. This direction of research, i.e. focusing on a certain property
of the real rank and examining where and when it breaks down for the binary or boolean rank, helps to clarify the differences
and similarities between these rank functions and hopefully will lead to an example with a stronger separation.

The same technique as used in the proof of Theorems 3 and 4,
yields a similar gap between the non-negative rank of $A$ and the real rank of $A$.
That is, for every $d$, there exists a binary matrix $A$ such that $R_{\mathbb{R}^+}(A) = 4d$ and $R_{\mathbb{R}}(A) = 3d$
(See remark after the proof of Theorem~4 in Subsection~\ref{Gap Section boolean}).
We also note that Watson showed recently in~\cite{Watson} a family of binary matrices in which the non-negative rank is $d$ and
the binary rank is $4d/3$, for every $d$ divisible by $9$.

Finally, even if a given matrix $A$ does not have the Augmentation property for the binary/boolen rank,
one might still hope that the rank cannot increase by too much when augmenting $A$ simultaneously
with a subset of vectors that do not increase its rank when it is augmented with each one separately.
We prove that this is not the case, and that it is possible to make the rank arbitrarily large
by augmenting $A$ with such a subset of vectors.

\medskip\noindent {\bf Theorem 5:} {\em
For any $k$, there exists a matrix $A_k$ and vectors $x_1,...,x_{k}$,
such that  $R_{binary}(A_k|x_i) =  R_{binary}(A_k) = 4$ for $1 \leq x_i \leq k$, but
$R_{binary}(A_k|x_1,...,x_{k}) =  k+3$.
}

\medskip\noindent {\bf Theorem 6:} {\em
For any $k$, there exists a matrix $A_k$ and vectors $x_1,...,x_{k}$,
such that  $R_{bool}(A_k|x_i) =  R_{bool}(A_k) = 4$ for $1 \leq x_i \leq k$, but
$R_{bool}(A_k|x_1,...,x_{k}) \geq k$.
}\\

Thus, without additional assumptions, the Augmentation property breaks down in the worst
possible way, for both the binary and the boolean rank.

\medskip\noindent {\bf Organization:}
In Section~\ref{Binary Section} we give the formal definitions needed throughout the paper,
define the base graph and prove Theorem~1.
Section~\ref{section characterize} is devoted to proving Theorem~2,
and Section~\ref{gaps} to the proof of Theorems~3 and ~4.
Finally, Theorems 5 and 6 will be proved in Section~\ref{section increase rank}.
We conclude with some open problems in Section~\ref{section open problems}.

\section{The Augmentation Property}
\label{Binary Section}



As stated above, as opposed to the real rank, the Augmentation property does
not always hold for the binary and real rank.
We start by giving simple examples of binary matrices that illustrate this.
Recall that the binary/boolean rank of a binary matrix $A$ is equal to the size of a minimal partition/cover of the $1$'s of
$A$ by monochromatic rectangles (see Equation~\ref{binary boolean}).
Therefore, we will use these concepts interchangeably throughout the paper.

\BL
\label{bad matrix example}
There exists a binary matrix $A$ and two column vectors $x,y$ such that
$R_{binary}(A|x) = R_{binary}(A|y) = R_{binary}(A)$, but $R_{binary}(A|x,y) > R_{binary}(A)$.
\EL
\BPF
Consider the following matrix $A$ and the vectors $x,y$ that are added
respectively to the following two minimal partitions of $A$:
$$
(A|x) =
\left(
     \begin{array}{ccc|c}
 0 &\cellcolor{cyan}1 & 0&\cellcolor{cyan}1\\
 \cellcolor{red}1  & \cellcolor{red}1 & 0&0\\
 \cellcolor{red}1  & \cellcolor{red}1 & \cellcolor{yellow}1&0\\
 0 &  \cellcolor{cyan}1 & \cellcolor{yellow}1&\cellcolor{cyan}1\\
\end{array}
\right)
\qquad\qquad
(A|y) =
\left(
\begin{array}{ccc|c}
 0  & \cellcolor{cyan}1 & 0&\cellcolor{cyan}1\\
 \cellcolor{red}1  & \cellcolor{cyan}1 & 0&\cellcolor{cyan}1\\
 \cellcolor{red}1 & \cellcolor{yellow}1 & \cellcolor{yellow}1&0\\
0  & \cellcolor{yellow}1 & \cellcolor{yellow}1&0\\
\end{array}
\right)
$$
In both cases the binary rank remains $3$ after augmenting $A$ with $x$ or $y$ separately.
However, when we augment $A$ with $x$ and $y$ simultaneously, it is easy to verify that the binary rank increases to $4$.
One such possible minimal partition for the new matrix is presented:
$$
(A|x,y) =
\left(
     \begin{array}{ccc|cc}
 0  & \cellcolor{cyan}1 & 0&\cellcolor{cyan}1&\cellcolor{cyan}1\\
 \cellcolor{red}1  & \cellcolor{red}1 & 0&0&\cellcolor{red}1\\
 \cellcolor{green}1  & \cellcolor{green}1 &\cellcolor{green}1&0&0\\
 0  & \cellcolor{yellow}1 & \cellcolor{yellow}1&\cellcolor{yellow}1&0\\
\end{array}
\right)
$$

The reason for the increase in the binary rank when augmenting $A$ with both $x$ and $y$ simultaneously,
is that $A$ has two different minimal partitions where $x$ and $y$ join different
rectangles in each one of these partitions, but $A$ does not have a minimal partition
to which both $x$ and $y$ can join.
\EPF


\BL
\label{boolean non aug}
There exists a binary matrix $A$ and two column vectors $x,y$,
such that $R_{bool}(A|x) = R_{bool}(A|y) = R_{bool}(A)$
but $R_{bool}(A|x,y) > R_{bool}(A)$.
\EL
\BPF
Consider the following matrix $A$ after augmenting it with the vectors $x,y$ separately,
and the two covers of the $1$'s of the augmented matrices. The orange cells belong both to the red
rectangle and to the yellow rectangle in the cover of the $1$'s in $(A|x)$:
$$
(A|x) =
\left(
     \begin{array}{ccc|c}
 \cellcolor{red}1 & \cellcolor{red}1 & \cellcolor{red}1 &0\\
 \cellcolor{red}1 & \cellcolor{orange}1 & \cellcolor{orange}1 &\cellcolor{yellow}1\\
 0 & \cellcolor{yellow}1 & \cellcolor{yellow}1 &\cellcolor{yellow}1\\
 0 & 0 & \cellcolor{green}1  &0\\
\end{array}
\right)
\qquad\qquad
(A|y) =
\left(
     \begin{array}{ccc|c}
  \cellcolor{red}1 & \cellcolor{red}1 & \cellcolor{red}1 &0\\
 \cellcolor{red}1 & \cellcolor{red}1 & \cellcolor{red}1&0\\
 0 & \cellcolor{yellow}1 & \cellcolor{yellow}1 &\cellcolor{yellow}1\\
 0 & 0& \cellcolor{green}1  &0\\
 \end{array}
 \right)
$$
In both cases the boolean rank of $A$ remains $3$, even after augmenting it with $x$ or $y$ separately.
However, when we augment $A$ with  $x$ and $y$ simultaneously, it is easy to verify that the boolean rank
increases to 4. One such possible minimal cover for the augmented matrix is presented:
$$
(A|x,y)=
\left(
     \begin{array}{ccc|cc}
 \cellcolor{red}1 & \cellcolor{red}1 & \cellcolor{red}1 &0&0\\
 \cellcolor{cyan}1 & \cellcolor{cyan}1 & \cellcolor{cyan}1 &\cellcolor{cyan}1&0\\
 0 & \cellcolor{yellow}1 & \cellcolor{yellow}1 &\cellcolor{yellow}1&\cellcolor{yellow}1\\
 0 & 0& \cellcolor{green}1  &0&0\\
\end{array}
\right)
$$
\EPF\\

It is interesting to note that although the matrix $A$ given in the proof of Lemma~\ref{boolean non aug}
does not have the augmentation property for the boolean rank, it does have it for the binary rank.
In fact, using the concepts that were mentioned briefly in the introduction and will be
defined in detail in the next subsections, we will see that
$A$ has a disjoint in rows base with respect to the binary rank, and it also has the
Unique base rows sums property for the binary rank. Hence, as we will prove shortly,
$A$ is an example of a matrix whose base graph with respect to the binary rank has one source,
whereas the base graph of $A$ under the boolean rank has at least two sources.

\subsection{Bases and Spanning Sets for Binary Matrices}
\label{Bases section}

In the following sub-sections we will prove that some matrices have a unique minimal
partition/cover such that each other minimal partition/cover can be obtained from it.
As we will prove, this is a sufficient and necessary condition for the Augmentation property to hold
with respect to the binary/boolean rank.
We first need to define the concept of a spanning set and a base for the columns of a matrix $A$:

\BD[Base, Spanning Set]
A set of vectors $X$ {\em spans} a set of vectors $Y$ with respect to $(S,+_S,\cdot_S)$,
if the elements of the vectors in $X$ are from $S$, and every vector in $Y$ can be
represented as a linear combination of vectors from $X$  with coefficients from the set $S$,
where the operations are $+_S,\cdot_S$.

If the set $Y$ is the set of columns of a matrix $A$,
then we say that $X$ is a {\em Spanning Set}
for the columns of $A$ with respect to $(S,+_S,\cdot_S)$.
A minimal spanning set for the columns of $A$ is called
a {\em Base} for the columns of $A$ with respect to $(S,+_S,\cdot_S)$,
and the {\em size} of the base is the number of vectors in it.

When the spanning set is over $(binary,+,\cdot)$, it is called a {\em Binary Spanning Set}
and the base is called a {\em Binary Base} for the columns of $A$,
and when discussing $(bool,+,\cdot)$ the spanning set is called a
{\em Boolean Spanning Set} and the base a {\em Boolean Base}.
When the underlying set $(S,+_S,\cdot_S)$ is clear from the context, we simply
write base or spanning set, instead of binary/boolean base/spanning set.
\ED

The following proof is similar to a proof that appears in Cohen and Rothblum for
the nonnegative rank~\cite{Cohen}, and we include it for completeness.
It shows that in fact the columns of $U$ in an optimal binary/boolean decomposition
$U\cdot V = A$, are a binary/boolean base for the columns of $A$.

\BL
Let $A$ be a binary matrix of size $n\times m$. Then:
\begin{itemize}
\item
 $R_{binary}(A) = d$
if and only if there exists a binary base of size $d$ for the columns of $A$.
\item
 $R_{bool}(A) = d$
if and only if there exists a boolean base of size $d$ for the columns of $A$.
\end{itemize}
\EL
\BPF
We prove the lemma for the binary rank. The proof for the boolean rank is similar.
Assume first that $R_{binary}(A) = d$.
Then $A$ has an optimal binary decomposition $A = U\cdot V$, where $U$ is an $n\times d$ binary matrix
and $V$ a $d \times m$ binary matrix.
Thus for every $i$ and $j$, it holds that $a_{i,j} = u_{i,1} v_{1,j}+u_{i,2}v_{2,j}+...+u_{i,d}v_{d,j}$.
Therefore, for every $j$, the $j$'th column of $A$ is a linear combination of the columns
$u_1,...,u_d$ of $U$ with  $0,1$ coefficients. Since $d$ is minimal then $u_1,...,u_d$ is a base of $A$.

Assume now that  $u_1,...u_d$ is a binary base for $A$. Thus for every $j$, the $j$'th  column of $A$
is a linear combination of  $u_1,...u_d$ with coefficients in $\{0,1\}$.
Define $U$ to be the $n \times d$ binary matrix whose columns are  $u_1,...u_d$
and let $V$ be the $d \times m$ binary matrix, where the $j$'th column of $V$ contains the coefficients
in the linear combination of $u_1,...u_d$ that result in the $j$'th column of $A$.
Therefore, $A = U\cdot V$. From the minimality of the base follows the minimality of the number of columns of $U$,
and thus  $U\cdot V$ is an optimal binary decomposition of $A$, and the binary rank of $A$ is $d$ as required.
\EPF\\

In fact, the vectors of each such binary/boolean base $U$, correspond to the rows
of a monochromatic rectangle of $1$'s
in a minimal partition/cover of $A$ as follows:
a  vector $u\in U$ has $1$'s in the entries corresponding to the rows of a rectangle
in the partition/cover and zeros in all other entries.

\begin{exmp}
\label{2 sources example}
{\em
The matrix $A$ given in the proof of Lemma~\ref{bad matrix example},
has three possible bases, $U_1,U_2,U_3$ for the binary rank,
that correspond to three different minimal partitions of this matrix:
$$
A = \left(
     \begin{array}{ccc}
 0 &  \cellcolor{cyan}1 & 0\\
  \cellcolor{red}1  & \cellcolor{red}1 & 0\\
 \cellcolor{red}1  & \cellcolor{red}1 & \cellcolor{yellow}1\\
  0 &  \cellcolor{cyan}1 & \cellcolor{yellow}1\\
\end{array}
\right)\qquad \qquad
U_1 = \left(
     \begin{array}{c}
       0 \\
       1 \\
      1\\
       0 \\
     \end{array}
   \right),
\left(
     \begin{array}{c}
       1 \\
       0 \\
      0\\
       1 \\
     \end{array}
   \right),
   \left(
     \begin{array}{c}
       0 \\
       0 \\
      1\\
       1 \\
     \end{array}
   \right)
$$
$$
A = \left(
     \begin{array}{ccc}
 0  & \cellcolor{cyan}1 & 0\\
 \cellcolor{red}1  & \cellcolor{cyan}1 & 0\\
 \cellcolor{red}1  & \cellcolor{yellow}1 & \cellcolor{yellow}1\\
 0  & \cellcolor{yellow}1 & \cellcolor{yellow}1\\
\end{array}
\right)\qquad \qquad
U_2 = \left(
     \begin{array}{c}
       0 \\
       1 \\
      1\\
       0 \\
     \end{array}
   \right),
\left(
     \begin{array}{c}
       1 \\
       1 \\
      0\\
       0 \\
     \end{array}
   \right),
   \left(
     \begin{array}{c}
       0 \\
       0 \\
      1\\
       1 \\
     \end{array}
   \right)
$$

$$
A = \left(
     \begin{array}{ccc}
 0  & \cellcolor{cyan}1 & 0\\
 \cellcolor{red}1  & \cellcolor{cyan}1 & 0\\
 \cellcolor{red}1  & \cellcolor{cyan}1 & \cellcolor{yellow}1\\
 0  & \cellcolor{cyan}1 & \cellcolor{yellow}1\\
\end{array}
\right)\qquad \qquad
U_3 = \left(
     \begin{array}{c}
       0 \\
       1 \\
      1\\
       0 \\
     \end{array}
   \right),
\left(
     \begin{array}{c}
       1 \\
       1 \\
      1\\
       1 \\
     \end{array}
   \right),
   \left(
     \begin{array}{c}
       0 \\
       0 \\
      1\\
       1 \\
     \end{array}
   \right)
$$
}
\end{exmp}

Recall that in Lemma~\ref{bad matrix example}, we showed for the above matrix $A$, that there
exist two vectors $x,y$, such that $R_{binary}(A|x) = R_{binary}(A|y)= R_{binary}(A)$,
but when augmenting the matrix with both vectors simultaneously we get that $R_{binary}(A|x,y)> R_{binary}(A)$.
Note that in each one of the bases presented above,
some of the vectors in the base intersect in rows, that is have $1$'s in the same row.
Also note that in this case, although the bases $U_1$ and $U_2$ both span the base $U_3$,
the matrix $A$ does not have a base that spans all other bases of $A$, with respect to the binary rank.
As we prove later, these two facts are not a coincidence and they
are the reason that we could not augment the matrix $A$ with both vectors simultaneously
without increasing its binary rank.

We first show that if we can augment a matrix with a vector without increasing its
rank, then there must exist a base for the matrix that spans this vector.
This property was also evident in the example given in Lemma~\ref{bad matrix example},
where each one of the vectors added to the matrix "joined" an existing minimal partition of the matrix.

\BCM
\label{vector spanned by base}
Let $A$ be an $n\times m$ binary matrix, let $x$ be a column vector and let $R_S()$ be a rank function
over $(S,+_S,\cdot_S)$.
Then $R_S(A|x) = R_S(A)$ if and only if there exists a base $U$ of $A$ such that $U$ spans $x$
with respect to $(S,+_S,\cdot_S)$.
\ECM
\BPF
Assume first that $R_S(A|x) =  R_S(A) = k$.
Hence, there exists an optimal decomposition of the matrix $(A|x)$ such that
$(A|x) = U \cdot V$, where $U$ is an $n \times k$ matrix and $V$ a $k \times (m+1)$ matrix,
the elements of $U,V$ are from $S$ and the addition and multiplication are $+_S,\cdot_S$.
But in this case we also have that $A = U \cdot V'$, where $V'$ is the matrix obtained from $V$ by removing
its last column. Hence, the columns of $U$ are a base of $A$ that spans $x$ with respect to $(S,+_S,\cdot_S)$.

The other direction of the Claim is trivial.
\EPF

\subsection{The Base Graph of a Matrix}
\label{base graph subsection}
We define the {\bf Base Graph}, $G_A$, of a given binary matrix $A$ with respect to $(S,+_S,\cdot_S)$,
as a directed graph that has a vertex $U$ for each base $U$ of $A$,
and has a directed edge from $U$ to $V$ if $U$ spans $V$.
A vertex in this graph is called a {\bf source} if there are no edges entering this vertex.

For example, the Base graph of the matrix $A$ presented in Example~\ref{2 sources example}
has three vertices $U_1,U_2,U_3$, and the edges are $(U_1,U_3),(U_2,U_3)$.
The vertices $U_1$ and $U_2$ are both sources of the graph.

\BD
The base graph is called {\em transitive} if for every three vertices $X,Y,Z$ in the graph,
if $(X,Y)$ and $(Y,Z)$ are edges in the graph, then also $(X,Z)$ is an edge in the graph.
\ED


As the following claim proves, the base graph is always transitive
with respect to $(binary, +,\cdot)$ or $(bool, +,\cdot)$.

\BCM
\label{transitive}
Let $X,Y,Z$ be three subsets of binary vectors. If $X$ spans $Y$ and $Y$ spans $Z$,
then also $X$ spans $Z$, with respect to $(binary, +,\cdot)$ or $(bool, +,\cdot)$.
\ECM
\BPF
We first prove the claim for the binary case. Assume that $X$ spans $Y$ and $Y$ spans $Z$,
and let $z \in Z$. Since $Y$ spans $Z$, there exist $y_1,...,y_k \in Y$ such that $z = y_1+...+y_k$.
But  $X$ spans $Y$ and so there exist $x_{1,1},...,x_{1,j_1},x_{2,1},...,x_{2,j_2},...,x_{k,1},...,x_{k,j_k}\in X$
such that:
\begin{equation}
\label{sum for y}
y_1 = (x_{1,1}+...+x_{1,j_1}), \ y_2 = (x_{2,1}+...+x_{2,j_2}),...,y_k=(x_{k,1}+...+x_{k,j_k}).
\end{equation}
Since all vectors are $0,1$ vectors, then $y_1,...,y_k$ are all disjoint
(that is, do not have $1$'s in the same entry),
and for the same reason also all $x_{i,j}$ in the above sums are disjoint,
and thus we get that:
\begin{equation}
\label{sum for z}
z = (x_{1,1}+...+x_{1,j_1})+(x_{2,1}+...+x_{2,j_2})+...+(x_{k,1}+...+x_{k,j_k}).
\end{equation}
Hence $X$ spans $Z$ with respect to $(binary, +, \cdot)$  as claimed.

As to the boolean case, the proof is identical up to and including  Equation~\ref{sum for y}.
Now since $1+1 = 1$ under the boolean algebra,
then we get Equation~\ref{sum for z} trivially.
Hence $X$ spans $Z$ with respect to $(bool, +, \cdot)$ as claimed.
\EPF\\

While for the real rank it is always true that every base of a given matrix $A$ spans all other bases
of $A$, this is not true for the binary or boolean rank of binary matrices. In fact, two bases can never span
each other:

\BCM
\label{one direction base}
For any two bases $U$ and $V$  of a binary matrix $A$, at most one of them spans the other
with respect to $(binary, +,\cdot)$ or $(bool, +,\cdot)$.
\ECM
\BPF
We first prove the claim for the binary case.
Assume by contradiction that $U$ spans $V$ and $V$ spans $U$, and let $u\in U$ be a vector such that $u \not \in V$.
By our assumption we can represent $u$ as a sum of vectors from $V$, $u = v_1 + ...+v_k$ for some $k$.
Also we can represent each $v_i$ in this sum as a sum of vectors from $U$, $v_i = u_{i,1}+...+u_{i,j_i}$.

If for each $v_i$, $1 \leq i \leq k$, it holds that $u$ is not one of the vectors that spans $v_i$, that is
$ u \neq u_{i,1},...,u_{i,j_i} $, then we have that:
\begin{equation}
\label{sum for u}
u =  (u_{1,1}+...+u_{1,j_1}) + ...+ (u_{k,1}+...+u_{k,j_k}).
\end{equation}
That is, $u$ is spanned by other vectors in $U$ and so we can remove $u$ from $U$ and get a smaller base,
in contradiction to the minimality of the base.

Otherwise, there exists an $i$ such that $u$ is one of the vectors that spans $v_i$.
Assume for simplicity that it is $v_1$ and that $u = u_{1,1}$. Then:
\begin{equation}
\label{sum2}
u =  (u + u_{1,2}+...+u_{1,j_1}) + v_2+...+v_k.
\end{equation}
But in this case, since all vectors are $0,1$ vectors, then it must be that
$u_{1,2},...,u_{1,j_1}, v_2,...,v_k$ are all the all $0$ vector.
Therefore, we get that $u = v_1$
in contradiction to our assumption that $u$ is a vector that does not belong to $V$.

The proof of the claim for the boolean case is again similar
and identical up to and including Equation~\ref{sum2}.
Now note that since all vectors are $0,1$ vectors and $1+1 = 1$ under the boolean algebra, then it must be the case
that $u_{1,2},...,u_{1,j_1}, v_2,...,v_k$ all have zeros where $u$ has zeros. In particular,
$v_1 = u + u_{1,2}+...+u_{1,j_1} = u$,
in contradiction to our assumption that $u$ is a vector that does not belong to $V$.
\EPF

\begin{coro}
\label{one spanning base}
A binary matrix $A$ can have at most one base that spans all other bases of $A$
with respect to $(binary, +,\cdot)$ or $(bool, +,\cdot)$.
\end{coro}

\begin{coro}
\label{acyclic}
The Base graph of a binary matrix $A$ is always acyclic
with respect to $(binary, +,\cdot)$ or $(bool, +,\cdot)$.
\end{coro}
\BPF
By Claim~\ref{transitive}, the Base graph is transitive for both $(binary, +,\cdot)$ and $(bool, +,\cdot)$.
Therefore, if there is a directed cycle in the graph, then there exist two bases $U,V$,
such that $U$ spans $V$ and $V$ spans $U$ (with respect to the given rank function), in contradiction to
Claim~\ref{one direction base}.
\EPF

\BCM
\label{singlesource}
The Base graph  $G_A$ has a single source if and only if the matrix $A$ has a (single) base
that spans all other bases, with respect to $(binary, +,\cdot)$ or $(bool, +,\cdot)$.
\ECM
\BPF
The proof for both the binary and the boolean case is identical.
We first show that if the Base graph has a single source $U$, then there is an edge in $G_A$
from $U$ to all other vertices, and thus $U$ spans all other bases.
Otherwise, let $V$ be a vertex such that there is no edge from $U$ to $V$.
By Claim~\ref{transitive} there is also no path from $U$ to $V$.
We walk backwards from $V$ along edges in the graph $G_A$, until we
either return to a vertex we were already at, in contradiction
to Corollary~\ref{acyclic}, or we can not go backwards since we arrive at a source,
in contradiction to the fact that $U$ is a single source.
Hence, indeed there is an edge from $U$ to all other vertices of $G_A$, and so the base $U$ spans
all other bases of $A$. Also since $U$ is a source, no other base spans $U$, and thus $U$ is the only
base that spans all other bases, as claimed.

On the other hand, if the matrix $A$ has a (single) base $U$ that spans all other bases,
then there are edges from $U$ to all other vertices of $G_A$. Thus, no other vertex can be a source.
Also, by Claim~\ref{one direction base}, all other bases do not span $U$, and thus there is no edge entering $U$.
Therefore, $U$ is a single source in $G_A$ as claimed.
\EPF

\subsection{A Sufficient and Necessary Condition for the Augmentation Property}
The following theorem proves that a necessary and sufficient condition for the Augmentation property
for the boolean or the binary rank, is having a base that spans all other bases of a given matrix,
or in other words, the Base graph of the matrix should have exactly one source.

\BT
\label{one spanning base theorem}
Let $A$ be a binary matrix such that $R(A) = k$,
where $R()$ is either the binary or boolean rank function.
\begin{itemize}
\item
If  $A$ has a base that spans all other bases of $A$,
then for every set of vectors  $x_1,...,x_t$ such that $R(A|x_i) = k$, for $1 \leq i \leq t$,
it holds also that  $R(A|x_1,...,x_t) = k$.
\item
If there is no base that spans all other bases of $A$, then
 there exists a subset of vectors $x_1,...,x_t$ such that $R(A|x_i) = k$, for $1\leq i \leq t$,
but  $R(A|x_1,...,x_t) > k$.
\end{itemize}
\ET
\BPF
The proof is identical for both the binary and boolean rank,
and follows from the claims and corollaries proved in Section~\ref{base graph subsection} for both rank functions.
Specifically, by Claim~\ref{singlesource}, the matrix $A$ has a (single) base that spans all other bases
if and only if the Base graph $G_A$ has a single source:
\begin{itemize}
\item
Assume that $A$ has a base $U$ that spans all other bases of $A$.
By Claim~\ref{vector spanned by base}, for every vector $x_i$ such that $R(A|x_i) = k$,
there must exist a base that spans $x_i$.
But $U$ spans all bases and so by Claim~\ref{transitive}, $U$ also spans $x_i$.
Therefore $U$ spans all vectors $x_1,...,x_t$, and so  $R(A|x_1,...,x_t) = k$.
\item
Otherwise, $G_A$ has at least two sources $U,V$.
It is clear that $R(A|U) = k$ and $R(A|V) = k$, but since
$U$ and $V$ are sources of the graph $G_A$, there is no base of $A$ that spans both $U$ and $V$.
Therefore, if we augment $A$ with the set of vectors in both bases $U,V$, we will get that
$R(A|U,V) > k$.
\end{itemize}
\EPF\\

\section{Characterizations of Matrices that have the Augmentation Property for the Binary Rank}
\label{section characterize}
Although Theorem~\ref{one spanning base theorem}
gives a sufficient and necessary condition for the Augmentation property,
it does not characterize the structure of the matrices
that have such a unique base that spans all other bases.
In the following subsections we show a characterization of binary matrices,
which guaranties that
a binary matrix has a base that spans all other bases of the matrix with respect to $(binary, +, \cdot)$,
and thus the Augmentation property holds for such matrices.

We note that most of this section deals with the binary rank,
besides a short discussion in Subsection~\ref{boolean rank excluded} that proves that this
characterization does not hold for the boolean rank. Therefore, for the sake of the simplicity of presentation,
in this section we just write: base, a base spans a vector, and so on,
and use the operations $+,\cdot$, with the understanding that they are all
with respect to $(binary,+,\cdot)$.

\subsection{Disjoint in Rows Bases for the Binary Rank}
\label{Section disjoint in rows}
The first simple observation is that a matrix that has a full row rank with respect to the binary rank, has the Augmentation
property:

\BL
\label{full row rank}
Let $A$ be an $n\times m$ binary matrix, where $R_{binary}(A) = n$. Then $A$ has the Augmentation property.
\EL
\BPF
Since $R_{binary}(A) = n$ then the columns of the identity matrix $I_n$ are a base for $A$.
But the identity matrix spans of course every base of $A$.
\EPF\\

The base $I_n$ is a special case of the following type of bases:

\BD[Disjoint in Rows Base]
A base is called {\em disjoint in rows} if every two vectors in the base do not share $1$'s in rows.
\ED

It is interesting to note that a disjoint in rows base is unique.

\BL
\label{unique base}
Let $A$ be a binary matrix of size $n\times m$.
Then $A$ has at most one base that is disjoint in rows.
\EL
\BPF
Assume that $A$ has a base that is disjoint in rows, and let $d = R_{binary}(A)$.
We will prove by induction on the number of rows of $A$ that it is unique.
We note that the binary rank of the all zero matrix is defined as $0$.

Assume for simplicity that $A$ does not have any zero rows or columns.
Otherwise, if for example the $i$'th row of $A$ is zero, then in any optimal binary decomposition $A= U\cdot V$,
the $i$'th row of $U$ must also be zero (and so all disjoint in rows bases are identical in this sense),
or we get an all zero row in $V$, in contradiction to the optimality of the decomposition.

The basis of the induction of a matrix $A$ with only one row is clear,
since there is only one base, and that is the vector $(1)$.

Now assume that $A$ has $n \geq 2$ rows, and let $U$ be
a disjoint in rows base for $A$. We will prove that $U$ is a unique disjoint in rows base for $A$.
Assume that $A$ has a column $v$ that is not all $1$. Otherwise, $A$ is the all one matrix,
and in this case there is only one base, which is the all one vector.

We can change the order of rows and columns of $A$ so that $v$ is the leftmost column of $A$
and all $1$'s of $v$ are at the first $i$ rows of $A$, where $i$ is the number of $1$'s that $v$ has.
Assume w.l.o.g. that  $u_1,...,u_k$ is the set of vectors in the base $U$ that span $v$,
that is $v =  u_1+...+ u_k$. Then it must be the case that the $n-i$ last entries of $u_1,...,u_k$ are all zeros.

Also since $u_1+...+u_k = v =  (1,1,...,1,0,0,0)^T$ and the base is disjoint in rows,
then the remaining vectors  $u_{k+1},...,u_d$ of the base $U$ must have only zeros in their first $i$ positions.
That is, $U$ is of the following form:
$$
U =
\left(
     \begin{array}{cc}
       X & 0 \\
       0 & Y
     \end{array}
   \right),
$$
where $X$ is an $i \times k$ binary matrix and $Y$ is an $(n-i) \times (d-k)$ binary matrix.

Denote by $B$ the submatrix of $A$ that is composed of the first $i$ rows of $A$,
and by $C$ the submatrix of $A$ that is composed of the last $n-i$ rows of $A$.
Obviously $X$ spans the columns of $B$ and $Y$ spans the columns of $C$.
Also $X$ must be a base for $B$, since otherwise we could replace $X$ with a base of
$B$ and get a base for $A$ of size smaller than $d$. Similarly $Y$ must also be a base for $C$.

Finally, both $X$ and $Y$ are disjoint in rows.
Thus, by our induction hypothesis, $X$ is the unique disjoint in rows base for $B$,
and $Y$ is the unique disjoint in rows base for $C$.
Therefore, $U$ must be the unique disjoint in rows base for $A$.
\EPF\\

The following claim is also easy to verify.
\BCM
Let $A$ be a binary matrix and let $U$ be a base of $A$.
Then any subset of vectors of  $U$
that span a certain column of $A$ must be disjoint in rows.
\ECM

\begin{exmp}
\label{example single source}
{\em
The first two partitions of the given matrix $A$ correspond to bases
 $U_1$ and $U_2$ that are not disjoint in rows.
$$
A = \left(
     \begin{array}{ccc}
  \cellcolor{green}1 & \cellcolor{red}1 &  0 \\
  0 & \cellcolor{red}1  & \cellcolor{cyan}1  \\
 0 & \cellcolor{red}1  & \cellcolor{cyan}1  \\
 0 & 0  & \cellcolor{cyan}1 \\
\end{array}
\right)\qquad\qquad
U_1 = \left(
     \begin{array}{c}
       1 \\
       0 \\
      0\\
       0 \\
     \end{array}
   \right),
\left(
     \begin{array}{c}
       1 \\
       1 \\
      1\\
       0 \\
     \end{array}
   \right),
   \left(
     \begin{array}{c}
       0 \\
       1 \\
      1\\
       1 \\
     \end{array}
   \right)
$$

$$
A = \left(
     \begin{array}{ccc}
  \cellcolor{green}1 & \cellcolor{green}1 &  0 \\
  0 & \cellcolor{red}1  & \cellcolor{cyan}1  \\
 0 & \cellcolor{red}1  & \cellcolor{cyan}1  \\
 0 & 0  & \cellcolor{cyan}1 \\
\end{array}
\right)\qquad\qquad
U_2 = \left(
     \begin{array}{c}
       1 \\
       0 \\
      0\\
       0 \\
     \end{array}
   \right),
\left(
     \begin{array}{c}
       0 \\
       1 \\
      1\\
       0 \\
     \end{array}
   \right),
   \left(
     \begin{array}{c}
       0 \\
       1 \\
      1\\
       1 \\
     \end{array}
   \right)
$$
The third partition of the same matrix $A$ corresponds to a base $U_3$ that is
disjoint in rows.
$$
A = \left(
     \begin{array}{cccc}
  \cellcolor{green}1 & \cellcolor{green}1 & 0  \\
  0 & \cellcolor{red}1 & \cellcolor{red}1  \\
 0 & \cellcolor{red}1  & \cellcolor{red}1 \\
 0 & 0  & \cellcolor{cyan}1  \\
\end{array}
\right)\qquad\qquad
U_3 = \left(
     \begin{array}{c}
       1 \\
       0 \\
      0\\
       0 \\
     \end{array}
   \right),
\left(
     \begin{array}{c}
       0 \\
       1 \\
      1\\
       0 \\
     \end{array}
   \right),
   \left(
     \begin{array}{c}
       0 \\
       0 \\
      0\\
       1 \\
     \end{array}
   \right)
$$
As can be easily verified, $U_3$ spans the other two bases of the matrix,
and thus by Theorem~\ref{one spanning base theorem}, this matrix has the Augmentation property for the binary rank.
}
\end{exmp}

In the above example the disjoint in rows base spans all other bases, and
in fact if there are no identical rows in the matrix this is always the case
and is a generalization of Lemma~\ref{full row rank}:

\BL
\label{spanning base}
Let $A$ be a binary matrix of size $n\times m$, and assume that $A$ has a base $u_1,...u_d$
that is disjoint in rows and that $A$ does not have identical rows.
Then $u_1,...u_d$ spans every other base of $A$.
\EL
\BPF
Assume for simplicity that $A$ does not have any zero rows or columns.
We will show that in this case under the assumptions of the Lemma, the matrix
$A$ has a full row rank and thus the lemma follows directly from Lemma~\ref{full row rank}.

Since  $u_1,...u_d$ is a base, we can represent $A$ as $A = U\cdot V$,
where $U$ is an $n \times d$ binary matrix whose columns
are $u_1,...u_d$.
Since the vectors $u_1,...u_d$ are disjoint in rows and we assumed that $A$
has no identical rows, then this implies that each $u_i$ has exactly one $1$, and therefore
$U$ is in fact a $d\times d$ matrix whose columns are a permutation of the identity matrix $I_d$.
Therefore $A$ has a full row rank.
\EPF\\

However, as the following example shows, when the matrix $A$ has identical rows,
it is not always the case that a disjoint in rows base spans all other bases of $A$.

\begin{exmp}
\label{identical rows example}
{\em
Consider the following matrix $A$ of size $6\times 5$, where $R_{binary}(A)=5$,
and the two bases $U_1$ and $U_2$ of $A$. Note that $U_1$ is a disjoint in rows base
and it does not span $U_2$.
$$
A = \left(
     \begin{array}{ccccc}
\cellcolor{pink} 1& \cellcolor{pink}1&  \cellcolor{pink}1 & \cellcolor{pink}1 &\cellcolor{pink}1 \\
\cellcolor{pink} 1& \cellcolor{pink}1&  \cellcolor{pink}1 & \cellcolor{pink}1 &\cellcolor{pink}1 \\
0& 0& 0 & \cellcolor{red}1 &  \cellcolor{red}1  \\
0& \cellcolor{green}1 &\cellcolor{green}1 &0 &0 \\
0&  \cellcolor{cyan}1& 0 & 0 &  \cellcolor{cyan}1 \\
0&  0&  \cellcolor{yellow}1 & \cellcolor{yellow}1 &  0 \\
\end{array}
\right)\qquad\qquad
U_1 = \left(
     \begin{array}{c}
       1 \\
       1 \\
      0\\
       0 \\
      0\\
      0\\
      \end{array}
   \right),
 \left(
     \begin{array}{c}
       0 \\
       0 \\
      1\\
       0 \\
      0\\
      0\\
      \end{array}
   \right),
    \left(
     \begin{array}{c}
       0 \\
       0 \\
      0\\
       1 \\
      0\\
      0\\
      \end{array}
   \right),
     \left(
     \begin{array}{c}
       0 \\
       0 \\
      0\\
       0 \\
      1\\
      0\\
     \end{array}
       \right),
           \left(
     \begin{array}{c}
       0 \\
       0 \\
      0\\
       0 \\
      0\\
      1\\
     \end{array}
       \right)
$$

$$
A = \left(
     \begin{array}{ccccc}
\cellcolor{pink}1& \cellcolor{cyan}1&  \cellcolor{yellow}1 & \cellcolor{yellow}1 & \cellcolor{cyan}1 \\
\cellcolor{pink}1&\cellcolor{green} 1& \cellcolor{green}1&  \cellcolor{red}1 & \cellcolor{red}1 \\
0& 0&  0 & \cellcolor{red}1 &  \cellcolor{red}1 \\
0& \cellcolor{green}1 &\cellcolor{green}1 &0 &0 \\
0&\cellcolor{cyan} 1& 0 & 0 &  \cellcolor{cyan}1 \\
0& 0&  \cellcolor{yellow}1 & \cellcolor{yellow}1 &  0 \\
\end{array}
\right)\qquad\qquad
U_2 = \left(
     \begin{array}{c}
       1 \\
       1\\
       0 \\
      0\\
      0\\
      0\\
     \end{array}
   \right),
 \left(
     \begin{array}{c}
       1 \\
       0 \\
       0\\
       0 \\
      1\\
      0\\
     \end{array}
   \right),
    \left(
     \begin{array}{c}
       1 \\
       0 \\
      0\\
       0 \\
      0\\
      1\\
     \end{array}
   \right),
     \left(
     \begin{array}{c}
       0 \\
       1 \\
      0\\
       1 \\
      0\\
      0\\
     \end{array}
   \right),
   \left(
     \begin{array}{c}
       0 \\
       1 \\
      1\\
       0 \\
      0\\
      0\\
     \end{array}
   \right),
$$
}
\end{exmp}

The reason that the disjoint in rows base $U_1$ does not span the base $U_2$, is that in the second
partition of $A$ induced by $U_2$, there are $1$'s in the same column of the two identical rows of $A$,
that do not  belong to the same rectangle in this minimal partition,
whereas the two identical rows belong to the same rectangle in the first partition induced by $U_1$.

This can be seen also in the two optimal binary decompositions of $A$ presented below,
that correspond to the bases $U_1$ and $U_2$ respectively.
The  matrix $V_2$ that represents the rows of rectangles in the second partition,
contains  two subsets of rows whose sums are identical to the first row of $V_1$.
That is, the sums of rows $1,2,3$ and rows $1,4,5$ in $V_2$, are both equal to $(1,1,1,1,1)$,
which is the first row of $V_1$, and
represents the identical rows that belong to the same rectangle
in the first partition induced by the disjoint in rows base $U_1$.

$$A = U_1\cdot V_1 = \left(
     \begin{array}{ccccc}
       1 &0 &0&0&0\\
       1&0 &0&0&0\\
       0&1&0&0&0 \\
      0&0&1&0&0\\
      0&0&0&1&0\\
      0&0&0&0&1\\
     \end{array}
   \right)\cdot
 \left(
     \begin{array}{ccccc}
 1 & 1 &1&1&1\\
  0&0 &0&1&1\\
  0&1&1&0 &0\\
  0&1&0&0&1\\
  0&0&1&1&0\\
     \end{array}
   \right)
$$

$$A = U_2\cdot V_2 = \left(
    \begin{array}{ccccc}
       1 &1 &1&0&0\\
       1&0 &0&1&1\\
       0&0&0&0&1 \\
      0&0&0&1&0\\
      0&1&0&0&0\\
      0&0&1&0&0\\
     \end{array}
   \right)\cdot
  \left(
  \begin{array}{ccccc}
 1 & 0 &   0&   0&0\\
 0 & 1 &  0&  0&1\\
  0&  0&  1&  1&0\\
   0&  1 &1&0&0\\
 0&  0& 0& 1&1 \\
 \end{array}
   \right)
$$

\subsection{Matrices with Rows that are Sums of other Rows for the Binary Rank}
\label{section sums of rows}
As we will prove shortly, for matrices $A$ with identical rows that posses the
{\em Unique base rows sums Property} defined below,
it will always be true that a disjoint in rows base, if exists, spans all other bases, and thus such matrices
have the Augmentation property for the binary rank.
In fact we will prove a more general result that will allow other linear dependencies with $0,1$ coefficients
among the rows of $A$ (and not only identical rows).

\begin{exmp}
{\em
Consider the following matrix $A$, in which
the fourth row of $A$ is the sum of the first two rows of $A$ (that is, it is a linear combination of these rows).
$$
A = \left(
\begin{array}{cccc}
1  & 1 & 0\\
0  & 0 & 1\\
  0 &  1 & 1\\
  1 & 1 & 1\\
   \end{array}
    \right)
= U\cdot V = \left(
     \begin{array}{ccc}
       1 & 0 &0\\
      0&1&0\\
  0&0&1 \\
   1&1 &0\\
     \end{array}
   \right)\cdot
 \left(
     \begin{array}{ccc}
      1 & 1 &   0\\
         0&   0 &1\\
        0&  1&  1 \\
      \end{array}
   \right)
$$
Note that $A$ has an optimal decomposition $A = U\cdot V$, in which the rows of $V$ are all rows of $A$.
Furthermore, the coefficients of the linear combinations (sums) of the rows in $A$,
are found in the rows of the base $U$. Thus for example, the fourth row $a_4$ of $A$ can be computed as follows:
$$a_4 = u_4 \cdot V = u_{4,1}\cdot v_1 + u_{4,2}\cdot v_2 + u_{4,3}\cdot v_3 = v_1 + v_2,$$
where $u_4$ is the forth row of $U$, and $v_1,v_2,v_3$ are the rows of $V$. In fact, we can view the
rows of $U$ as a binary "code" which tells us which rows of $V$ should be summed to get a row in $A$.
}
\end{exmp}

\BCM
\label{Y in A}
Let $A$ be a binary matrix that has an optimal binary decomposition $A = U\cdot V$,
such that the rows of $V$ are rows of the matrix $A$.
Then all rows of $A$ are linear combinations of the rows in $V$ with $0,1$ coefficients.
\ECM
\BPF
Assume without loss of generality that the rows of $V$ are
the first $k$ rows of $A$. Each of the other rows of $A$ is a linear combination
of the rows of $V$ with $0,1$ coefficients. The rows of the
matrix $U$ are the vectors of coefficients of these linear combinations.
In particular, the first $k$ rows of $U$ form the identity matrix.
\EPF\\

Note that a matrix that has a disjoint in rows base always has such a decomposition as described in Claim~\ref{Y in A}.

\BCM
\label{full_rank}
Let $A$ be a binary matrix and $A = U\cdot V$ an
optimal binary decomposition of $A$.
Then $U$ and $V$ are full binary rank matrices.
\ECM
\BPF
Assume that $A$ is an $n\times m$ matrix and that $R_{binary}(A) = k$.
Therefore  $U$ is an $n\times k$ matrix and $V$ is a $k \times m$ matrix.
Assume first by contradiction that $V$ is not a full rank matrix. Therefore $V$ has
an optimal binary decomposition $V = X\cdot Y$, such that $X$ is a $k \times t$ binary matrix and $Y$ is a
$t \times m$ binary matrix for some $t < k$. Therefore, we can express $A$ as
$$A = U \cdot (X \cdot Y) = (U \cdot X) \cdot Y = D \cdot Y,$$
where  $D$ is a matrix of dimension $n \times t$.

If $D$ is a binary matrix, then we get a binary decomposition of $A$ of size $t < k$, in contradiction to the fact that
$U\cdot V$ was an optimal binary decomposition of $A$.

Otherwise, $D$ is not a binary matrix and thus there exists an element $d_{i,j} \in D$, such that  $d_{i,j} \ne 0,1$.
Note that $D$ is a nonnegative matrix, since $U$ and $X$ are binary matrices.
Therefore, since $Y$ and $A$ are binary matrices,
this means that $i$'th row of $Y$ must be the all $0$ vector.
But this is in contradiction to the fact that $X\cdot Y$ is an optimal decomposition of $V$.

Thus, $V$ is a full rank matrix as claimed. In a similar way we can show that $U$ is a full rank matrix.
\EPF\\

The following lemma gives an alternative characterization of the type of matrices
that we study in this section.

\BL
\label{linear_dep}
A matrix $A$ has an optimal binary decomposition $A = U\cdot V$,
such that the rows of $V$ are rows of the matrix $A$,
if and only if after removing from $A$ rows that are linear combinations
with $0,1$  coefficients of other rows, the remaining matrix has a full binary rank.
\EL
\BPF
Assume first that $A$ has an optimal binary decomposition $A = U\cdot V$,
such that the rows of $V$ are the first (w.l.o.g.) rows of $A$.
By Claim~\ref{Y in A}, the remaining rows of $A$ are all linear combinations
of the first rows of $V$, with $0,1$ coefficients as defined by $U$.
Thus, if we remove them from $A$, then we get the matrix $V$, and by
Claim~\ref{full_rank} this matrix has a full binary rank.

Assume now that after removing from $A$ all rows that are linear combinations
with  $0,1$ coefficients of other rows, then the remaining $k$ rows of $A$ have a full binary rank.
Denote the remaining matrix by $V$, and define $U$ to be the matrix with the corresponding $0,1$ coefficients
of these linear combinations. It is clear that $A = U \cdot V$.

Finally, assume that the binary rank of $V$ is $k$.
Since $V$ is a subset of rows of $A$, then
$R_{binary}(A) \geq R_{binary}(V) = k$,
and since $A =U\cdot V$  is a binary decomposition of $A$, we also have that $R_{binary}(A) \leq k$.
Thus $R_{binary}(A)  = k$ and  $U\cdot V = A$ is an optimal binary decomposition of $A$.
\EPF\\

We now define a property which guarantees that a matrix that has such a decomposition as stated
in Lemma~\ref{linear_dep}, has the Augmentation property.

\BD[Unique base rows sums Property]
A matrix $A$ has the {\em Unique base rows sums} property if in any  optimal binary decomposition $A = X\cdot Y$ of $A$,
it holds that $Y$ does not contain two disjoint subsets of rows $y_{i_1},...,y_{i_s}$ and  $y_{j_1},...,y_{j_t}$,
such that $y_{i_1}+...+y_{i_s} = y_{j_1}+...+y_{j_t}$.
\ED

The following lemma proves that for binary matrices $A$ that have the Unique base rows sums property,
it holds that in any optimal binary decomposition $A = X\cdot Y$  of $A$, the matrices
$A$ and $X$ have the same linear dependencies among their rows.

\BL
\label{lem:irp}
Let $A$ be a binary matrix with the Unique base rows sums property,
and let $A=X\cdot Y$ be an optimal binary decomposition of $A$.
Then for any $t$ and any subset of indices $i_1,...,i_t$ of $t$ rows  in $X$ and $A$, we have
that  $x_{i_t} = \sum_{j=1}^{t-1} x_{i_j}$ if and only if $a_{i_t} = \sum_{j=1}^{t-1} a_{i_j}$.
\EL

\BPF
Assume without loss of generality that $i_1,...,i_t$ are the indices of the first $t$ rows of $A$ and $X$,
and denote by $A'$ and $X'$ the matrices that are composed of the first $t$ rows of $A$
and $X$ respectively. Thus of course $A' = X'\cdot Y$.

Let $v = (1,1,\ldots,1,-1) $ be a vector in dimension $t$.
If $x'_t = \sum_{j=1}^{t-1} x'_{j}$ then  $v^TX' = 0$,
which implies that $v^TA' = v^TX'\cdot Y = 0\cdot Y=0$.
Thus $a'_t = \sum_{j=1}^{n-1} a'_j$ as claimed.

On the other hand, assume that $x'_t \ne \sum_{j=1}^{t-1} x'_j$
and let $\alpha = \sum_{j=1}^{t-1} x'_j$ and $\beta = x'_t$.
Thus there is a coordinate $j$ such that $\alpha_j \ne \beta_j$.
Note also that $\sum_{j=1}^{t-1} a'_j = \alpha Y$ and $a'_t = \beta Y$.
We will prove that $\alpha Y \ne \beta Y$ and so $a'_t \ne \sum_{j=1}^{t-1} a'_t$ as claimed.
We separate the proof into two cases:
\begin{itemize}
\item
If $\alpha$ is not a binary vector then assume w.l.o.g. that $\alpha_1>1$.
In this case we claim that $\alpha Y$ must also not be a binary vector.
Otherwise, the first row of $Y$ must be the all zero vector,
in contradiction to the fact that $X\cdot Y$ is an optimal decomposition of $A$.
But $\beta Y = a'_t$ is a binary vector and so $\alpha Y \ne \beta Y$.
\item
Now assume that $\alpha$ is a binary vector,
and recall that $\alpha \ne \beta$. Hence $\alpha-\beta$ is a non-zero vector in $\{1,0,-1\}$.
Assume by contradiction that $\alpha Y = \beta Y$,
and so $(\alpha-\beta)Y = 0$. Let $k_1,...,k_s$ be the indices where $\alpha-\beta$ equals $1$, and let  $l_1,...,l_t$
be the indices where $\alpha-\beta$ equals $-1$. Then $y_{k_1}+...+y_{k_s} = y_{l_1}+...+y_{l_t}$, in contradiction
to our assumption that $A$ has the Unique base rows sums property.
\end{itemize}
\EPF

\BT
\label{7}
Let $A$ be binary matrix with an optimal binary decomposition $A = U \cdot V$,
such that the rows of $V$ are rows of the matrix $A$. If in addition, $A$
has the Unique base rows sums property, then the columns
of $U$ span every base of $A$, and thus $A$ has the Augmentation property.
\ET
\BPF
Let $k=R_{binary}(A)$, and assume without loss of generality that the rows of $V$ are
the first $k$ rows of $A$. By claim~\ref{Y in A},
each of the other rows of $A$ is a linear combination of the rows of $V$,
where the matrix $U$ contains the vectors of the $0,1$ coefficients
of these linear combinations.

Let $X\cdot Y=A$ be another optimal binary decomposition of $A$, and let $D$ form the first $k$ rows of $X$.
Since $A$ has the  Unique base rows sums property, then by Lemma~\ref{lem:irp},
the rows of $X$ and $A$ have the same linear dependencies.
Thus, since  $U$ contains the $0,1$ coefficients of the linear dependencies of the rows of $A$,
it also contains the $0,1$ coefficients of the linear dependencies of the rows of $X$.
Therefore, we have that $X=U\cdot D$, and thus $U$ spans $X$ as required.
\EPF

\begin{coro}
\label{coro unique sums}
Let $A$ be a binary matrix and let $U$ be a disjoint in rows base of $A$.
If, in addition, $A$ has the Unique base rows sums property, then $U$ spans every base of $A$,
and thus $A$ has the Augmentation property.
\end{coro}
\BPF
Since $U$ is a disjoint in rows base, there is a matrix $V$
whose rows are rows of the matrix $A$ such that
$A = U\cdot V$ is an optimal binary decomposition.
Thus the corollary follows directly from Theorem~\ref{7}.
\EPF\\

\subsection{Theorem~\ref{7}  does not hold for the Boolean Rank}
\label{boolean rank excluded}
We conclude this section by proving that
Theorem~\ref{7} and Corollary~\ref{coro unique sums} do not hold for the boolean rank.

\BL
There exists a binary matrix $A$ such that:
\begin{itemize}
\item
The matrix $A$ has the Unique base rows sums property.
\item
The matrix $A$ has a base that is disjoint in rows
(and thus $A$ has an optimal boolean  decomposition $A = U \cdot V$, where the rows of $V$ are rows of $A$).
\end{itemize}
However, $A$ does not have the Augmentation property with respect to the boolean rank.
\EL
\BPF
Let $A$ be the following binary matrix:
$$
A = \left(
     \begin{array}{ccc}
 1 & 1 & 1 \\
1 & 1 & 1 \\
 0 &1 & 1 \\
 0 & 0 & 1  \\
\end{array}
\right)
$$
As we saw in the proof of Lemma~\ref{boolean non aug},
 this matrix does not have the augmentation property for the boolean rank.

We first prove that in any optimal boolean decomposition $U\cdot V$ of $A$, with respect to the boolean rank,
the first column of $V$  has exactly one $1$ and two $0$'s. Consider an optimal boolean decomposition $A = U\cdot V$,
and assume by contradiction that the first column of $V$ has at least two $1$'s.
Assume w.l.o.g. that these two ones are at the first two positions of the column
(otherwise, substitute the order of the rows of $V$ and accordingly the order of the columns of $U$).
Denote by $u_i$ the $i$'th row of $U$ and by $v^j$ the $j$'th column of $V$.

Since $u_3 \cdot v^1 = a_{3,1} = 0$ and $u_4 \cdot v^1 = a_{4,1} = 0$,
then it must be that $u_{3,1} =  u_{3,2} = u_{4,1} = u_{4,2} = 0$.
We also have that $u_3 \cdot v^2 = a_{3,2} = 1$ and $u_3 \cdot v^3 = a_{3,3} = 1$.
Therefore it must be that $u_{3,3} = v_{3,2} = v_{3,3} = 1$.
Also since $u_4 \cdot v^3 = a_{4,3} = 1$, then $u_{4,3} = 1$.
Thus the decomposition looks as follows:
$$
 U\cdot V =
\left(
\begin{array}{ccc}
 u_{1,1} & u_{1,2} & u_{1,3}\\
 u_{2,1} & u_{2,2} & u_{2,3}\\
0 & 0 & 1\\
0 & 0 & 1\\
\end{array}
\right)\cdot
\left(
\begin{array}{ccc}
 1 & v_{1,2} & v_{1,3}\\
 1 & v_{2,2} & v_{2,3}\\
 v_{3,1} &1 &1\\
\end{array}
\right)
$$
But then we get a contradiction since $u_4 \cdot v^2 = 1 \neq 0 = a_{4,2}$.

Hence in any optimal boolean decomposition of $A$, the first column of $V$ has exactly one $1$.
Thus it is not possible for the sum of two rows of $V$ to be equal to the third row
(using the boolean algebra). Therefore, $A$ has the Unique base rows sums property
with respect to the boolean rank.

Furthermore, the following is an optimal boolean decomposition $U\cdot V$ of $A$,
such that the rows of $V$ are rows of the matrix $A$,
and $U$ is also a disjoint in rows base:
$$
A = U\cdot V =
\left(
\begin{array}{ccc}
 1 & 0 & 0\\
1  & 0 & 0\\
0  & 1 & 0\\
  0 &  0 & 1\\
\end{array}
\right)\cdot
\left(
\begin{array}{ccc}
 1 & 1 & 1\\
0  & 1 & 1\\
  0 &  0 & 1\\
\end{array}
\right).
$$
\EPF

\section{Creating a Gap between Different Rank functions}
\label{gaps}
In this section we will use the concept
of the base graph and the augmentation property defined in Section~\ref{Binary Section},
in order to design a technique that allows us to build a family of matrices
with a linear gap between their binary and real rank, and in a similar way a family of matrices
with a linear gap between their boolean and real rank.
Although this is not a large gap, we believe that the technique itself is interesting,
and hope that it can be used to find a family of matrices with a larger gap.

\subsection{A Family of Matrices with a Linear Gap between the Real and Binary Rank}
\label{Gap Section}
As proved in Section~\ref{Binary Section}, if the base graph of a given matrix $A$
has two sources with respect to the binary rank, we can increase the binary rank of $A$ by at least one,
by augmenting $A$ with the vectors that belong to the two bases that are represented by these two sources.

As we prove shortly, if the real rank of the original matrix $A$ equals its binary
rank, then the real rank of the augmented matrix does not increase, and
so we created a gap of size at least $1$ between the real and binary rank of a given matrix.

The question is if we can significantly amplify  this gap,
by creating a binary matrix $A$ whose base graph has many sources for the binary rank,
and then augmenting $A$ with the vectors in the bases represented by these sources.
If again the real rank of the augmented matrix does not increase,
then we get a larger gap between the real and binary rank.

We note that it is not always the case that if $R_{binary}(A|x)=R_{binary}(A)$
then also $R_{\mathbb{R}}(A|x)=R_{\mathbb{R}}(A)$ (see the discussion following Theorem~\ref{duplicate theorem}),
and thus the technique we describe shortly can not be used in such cases.

We first prove the following sufficient condition that guarantees that the real
rank of the augmented matrix does not increase, even when the binary rank of the augmented matrix increases:
\BCM
\label{real aug}
Let $A$ be a binary matrix where $R_{\mathbb{R}}(A) = R_{binary}(A)$.
Then if $x_1,...,x_t$ are vectors such that $R_{binary}(A|x_i)=R_{binary}(A)$ for $1\leq i \leq t$,
we have that $R_{\mathbb{R}}(A|x_1,...,x_t) =R_{\mathbb{R}}(A)$.
\ECM
\BPF
Since the real rank of any matrix is  bounded above by its binary rank, we have that
$R_{\mathbb{R}}(A) \leq R_{\mathbb{R}}(A|x_i)\leq R_{binary}(A|x_i)=R_{binary}(A) = R_{\mathbb{R}}(A)$,
for $1\leq i \leq t$. Hence, $R_{\mathbb{R}}(A|x_i) = R_{\mathbb{R}}(A)$ for $1\leq i \leq t$,
which implies that $x_i$ belongs to the subspace spanned by the columns of $A$.
Thus $R_{\mathbb{R}}(A|x_1,...,x_t) =R_{\mathbb{R}}(A)$.
\EPF\\

We now describe a simple construction that achieves the goal described above
and amplifies the gap between the real and binary rank:

Given a binary matrix $A$ of size $n\times m$,
we construct a new matrix $A'$ of size $n'\times m'$, where $n' = n\cdot d$ and $m' = m\cdot d$,
such that $A'$ has the matrix $A$ along its diagonal $d$ times,
and $0$ in every other position:

$$
A' = \left(
     \begin{array}{cccc}
A  &   0 &0 & 0 \\
  0  & A &0 & 0  \\
  0  & 0 & \ddots & 0    \\
 0 & 0 & 0 & A  \\
\end{array}
\right)
$$
In other words $A'=I_d \otimes A$, where $I_d$ is the $d \times d$ identity matrix
and $\otimes$ denotes the tensor product.

It is interesting to note that if $A$ has at least two bases,
then $A'$ has an exponential number of bases and sources, as the following two claims prove:
\BCM
If $A$ has $t$ bases then $A'=I_d \otimes A$ has at least $t^{d}$ bases (for the binary rank).
\ECM
\BPF
Let $U_1,...,U_t$ be the bases of $A$. Then for any choice of $d$ bases
$U_{i_1},\ldots,U_{i_d}$, the set $\{e_1\otimes U_{i_1}, \ldots , e_d \otimes U_{i_d}\}$
is a base of $A'$, where $e_1, \ldots, e_d$ are the standard base,
i.e. the columns of $I_d$.
There are $t^d$ ways to choose a base as above, and thus $I_d \otimes A$
has at least $t^d$ bases.
\EPF

\BCM
If the base graph of $A$ has at least $2$ sources,
then the base graph of $A'=I_d \otimes A$ has at least $2^{d}$ sources (for the binary rank).
\ECM
\BPF
As in the proof of the previous claim, observe that if
$\{U_{i_1},...,U_{i_{d}}\}$ are sources in the base graph of $A$,
then $\{e_1\otimes U_{i_1}, \ldots , e_d \otimes U_{i_d}\}$ is a source
in the base graph of $A'$. The rest of the proof is similar.
\EPF\\

We will need the following simple claim for the analysis of our construction:

\BCM
\label{diagonal}
Let $M$ be binary matrix of the following form:
$$
M = \left(
     \begin{array}{cc}
B  &   0 \\
  0  & C  \\
\end{array}
\right)
$$
where $B$ and $C$ are binary matrices. Then
$R_{binary}(M) = R_{binary}(B) + R_{binary}(C)$,
$R_{bool}(M) = R_{bool}(B) + R_{bool}(C)$
and $R_{\mathbb{R}}(M) = R_{\mathbb{R}}(B) + R_{\mathbb{R}}(C)$.
\ECM
\BPF
Let $U$ be a base of $B$ and $V$ be a base of $C$. Then, as we have noticed before
in a more restricted case, it holds that $\{e_1 \otimes U, e_2 \otimes V\}$ is a base
of M. The claim follows since the rank is equal to the size of a base, and the argument
holds similarly for the binary, boolean and real rank.
\EPF\\

The following claim follows easily from Claim~\ref{diagonal}:
\BCM
\label{diagonal coro}
For any binary matrix $A$ it holds that
$R_{binary}(I_d \otimes A) = d\cdot R_{binary}(A)$ and  $R_{\mathbb{R}}(I_d \otimes A) = d\cdot R_{\mathbb{R}}(A)$.
\ECM

We can now prove that if we augment $A'=I_d\otimes A$ with vectors that belong to bases that are sources in the base graph of $A'$,
then the binary rank of the resulting augmented matrix increases in comparison with the binary rank of $A'$:
\BL
\label{Augmented rank}
Let $A$ be a binary matrix whose base graph has (at least) $2$ sources $U_1$ and $U_2$.
Let $A'=I_d\otimes A$ and let $A''$ be the matrix that results from $A'$ by augmenting it as follows:
$$
A'' = \left(
     \begin{array}{cccc|ccccccccc}
A  &   0 &0 & 0 & U_1 & U_2 & 0 & 0 & 0 & 0 & 0 & 0 \\
  0  & A &0 & 0 & 0 & 0 & U_1 & U_2 &  0 & 0 & 0 & 0\\
  0  & 0 & \ddots & 0 & 0 & 0 & 0 & 0 & \ddots & \ddots & 0 & 0   \\
 0 & 0 & 0 & A &  0 & 0 & 0 & 0 & 0 & 0 & U_1 & U_2 & \\
\end{array}
\right)
$$
Then $R_{binary}(A'') \geq kd + d$, where $R_{binary}(A) = k$.
\EL
\BPF
Let $B=(A|U_1,U_2)$. Then, by changing the order of the columns of $A''$
we can get the matrix $I_d \otimes B$. Obviously $R_{binary}(A'')= R_{binary}(I_d \otimes B)$,
as permuting columns does not change the rank. It is therefore enough to prove the claim for $I_d \otimes B$.

Since $U_1$ and $U_2$ are sources of the base graph of $A$, then $R_{binary}(B) > R_{binary}(A)=k$.
By Claim~\ref{diagonal} and Claim~\ref{diagonal coro} we have that $R_{binary}(I_d \otimes B) = d\cdot R_{binary}(B)$.
Thus
$$
R_{binary}(I_d \otimes B) = d\cdot R_{binary}(B)\ge d(k+1) = dk+d.$$
\EPF\\

Consider for example the following matrix $A$ of size $4\times 3$,
whose binary rank is $3$ and whose base graph has two sources
(see Example~\ref{2 sources example}):

$$
A = \left(
     \begin{array}{cccc}
 0 & 1 & 0\\
1  & 1 & 0\\
1  & 1 & 1\\
  0 &  1 & 1\\
\end{array}
\right)
$$
Then the matrix $A'=I_d \otimes A$ is of dimension $(4 d) \times (3  d)$, has binary rank $3d$, and the number of
sources in the base graph of $A'$ is at least $2^{d}$.

Now let $A''$ be the matrix that results by augmenting $A'$ as in Lemma~\ref{Augmented rank}.
Then by Lemma~\ref{Augmented rank} we have that $R_{binary}(A'') \geq 3d + d = 4d$.

Note that the real rank of $A$ is also $3$, and $R_{\mathbb{R}}(A') = 3d = R_{binary}(A')$.
Therefore, by Claim~\ref{real aug}, the real rank of the augmented matrix $A''$ is $R_{\mathbb{R}}(A'') = R_{\mathbb{R}}(A') = 3d$.
We have thus proved:

\BT
\label{real binary gap}
For every $d$, there exists a matrix $M$
such that $R_{binary}(M) \geq 4d$ and $R_{\mathbb{R}}(M) = 3d$.
\ET

\subsection{A Summary of the Technique and Restrictions on $A$ for the Binary Rank}
As we saw, if we want to amplify the gap between the binary and real rank using the above
technique, we need a matrix $A$ that has the following properties:
\begin{enumerate}
\item  The base graph of $A$ has at least $2$ sources with respect to the binary rank.
\item
$R_{\mathbb{R}}(A) = R_{binary}(A)$, as this guarantees that the real rank will not increase when augmenting $A$.
\item
The binary rank $k$ of $A$ should be as small as possible,
so that we get the largest gap possible between the real and the binary rank of the augmented matrix $A''$.
\end{enumerate}

As the following claim proves, $k = 2$ is not possible, and
thus we need a matrix of binary rank at least $k = 3$, as was used
in the proof of Theorem~\ref{real binary gap}.

\BCM
Let $A$ be a binary matrix such that $R_{binary}(A) = 2$. Then the base graph of $A$ has only one source.
\ECM
\BPF
Note first that if $R_{binary}(A) = 2$ then in any optimal binary decomposition $X\cdot Y = A$ of $A$,
the matrix $Y$ has exactly two rows, and so the Unique base rows sums property holds trivially for $A$.
We show that in addition, $A$ has an optimal binary decomposition $U\cdot V = A$,
such that the rows of $V$ are rows of the matrix $A$,
and then it will follow from Theorem~\ref{7} that the matrix $A$ has one base that spans all other bases.

Assume that $A$ is an $n \times m$ matrix, and let $X\cdot Y = A$ be an optimal binary decomposition of $A$,
where $X$ is an $n \times 2$ binary matrix and $Y$ a $2 \times m$ binary matrix.
The possible rows of $X$ are of course $(0,1),(1,0),(1,1)$ and $(0,0)$.
Note that  $X$ must have at least two of these first three types of rows,
otherwise $X$ is not of full rank.

If $X$ has rows $(0,1)$ and $(1,0)$ then we are done, since in this case both rows of $Y$ are
rows of $A$. Otherwise, the rows of $X$ do not include either $(0,1)$ or $(1,0)$.
Assume it is the first case, and note that in this case $A$ has only two types of rows,
the first row of $Y$ and a row that is the sum of the two rows of $Y$ (and possibly rows that are all $0$).
Now define a binary matrix $V$ that has exactly these two rows that $A$ has,
and define a matrix $U$ whose rows are either $(1,0)$ or $(0,1)$ or $(0,0)$, according to the order of the rows in $A$.
\EPF\\

Furthermore, if $A$ does not have the unique rows sums property then Item $2$ in the summary
above does not hold.

\BCM
Let $A$ be an $n\times m$ binary matrix that does not have the unique rows sums property. Then $R_{\mathbb{R}}(A) < R_{binary}(A)$.
\ECM
\BPF
Assume that $R_{binary}(A)=k$. By our assumption, there exists an optimal binary decomposition $U\cdot V = A$ of $A$,
where $U$ is an $n\times k$ binary matrix, $V$ is a $k \times m$ binary matrix, and $V$ has two subsets of rows whose
sums are equal.

But this implies that $R_{\mathbb{R}}(V) <k$ and therefore $R_{\mathbb{R}}(A) <k$.
\EPF\\

We thus get the following restrictions on the matrix $A$ needed for this construction:
the matrix $A$ should have binary rank $k \geq 3$ and it should have the unique rows sums property.
Finally, $A$ can not have an optimal binary decomposition $U\cdot V = A$, such that the rows of $V$ are rows of $A$,
as otherwise by Theorem~\ref{7},  it will have the augmentation property,
and thus its base graph will have only one source.

\subsection{A Family of Matrices with a Linear Gap between the Real and Boolean Rank}
\label{Gap Section boolean}
As opposed to the binary rank, the boolean rank is not always larger than the real rank.
Still it is interesting to construct families of matrices in which there is a large gap between
the real and boolean rank.
Again by Theorem~\ref{one spanning base theorem}, we have to find a matrix $A$ whose
base graph under the boolean rank, has two sources, and then augment $A$ with the vectors that belong
to the two bases that are represented by these two sources. If the real rank of the augmented
matrix does not increase, then we have created a gap between the real and the boolean rank.

Consider again the matrix $A$ that was used in Section~\ref{Gap Section}:

\begin{equation}
\label{matrix for boolean gap}
A = \left(
     \begin{array}{cccc}
 0 & 1 & 0\\
1  & 1 & 0\\
1  & 1 & 1\\
  0 &  1 & 1\\
\end{array}
\right)
\end{equation}

It holds that $R_{bool}(A) = R_{\mathbb{R}}(A) = 3$.
Furthermore, the following two bases, $U_1$ and $U_2$, are
sources in the base graph of $A$ for the boolean rank:
$$
U_1 = \left(
     \begin{array}{c}
       0 \\
       1\\
       1 \\
      0\\
      \end{array}
   \right),
 \left(
     \begin{array}{c}
       1 \\
       0 \\
       0\\
       1 \\
      \end{array}
   \right),
    \left(
     \begin{array}{c}
       0 \\
       0 \\
      1\\
       1 \\
     \end{array}
   \right),\qquad\qquad
 U_2 = \left(
     \begin{array}{c}
       0 \\
       1\\
       1 \\
      0\\
      \end{array}
   \right),
 \left(
     \begin{array}{c}
       1 \\
      1 \\
       0\\
       0 \\
      \end{array}
   \right),
    \left(
     \begin{array}{c}
       0 \\
       0 \\
      1\\
       1 \\
     \end{array}
   \right)
 $$
Note that if we augment $A$ simultaneously with the vectors $x = (1,0,0,1)^T\in U_1$
and $y = (1,1,0,0)^T\in U_2$ then its boolean rank increases to $4$.

Recall that for this matrix we also have that $R_{\mathbb{R}}(A)= R_{binary}(A)$
and $R_{binary}(A|x) = R_{binary}(A|y) = 3$. Therefore, by Claim~\ref{real aug}
we have that $R_{\mathbb{R}}(A|x,y)= 3$. Thus we have that $R_{bool}(A|x,y) > R_{\mathbb{R}}(A|x,y)$.

We now amplify the gap by constructing the matrix $A'=I_d \otimes A$ that has $A$
on its diagonal $d$ times and augmenting it as follows:

$$
A'' = \left(
     \begin{array}{cccc|ccccccccc}
A  &   0 &0 & 0 & x & y & 0 & 0 & 0 & 0 & 0 & 0 \\
  0  & A &0 & 0 & 0 & 0 & x & y &  0 & 0 & 0 & 0\\
  0  & 0 & \ddots & 0 & 0 & 0 & 0 & 0 & \ddots & \ddots & 0 & 0   \\
 0 & 0 & 0 & A &  0 & 0 & 0 & 0 & 0 & 0 & x & y & \\
\end{array}
\right)
$$

Recall that in this case $R_{\mathbb{R}}(A') = 3d = R_{binary}(A')$, and so again
by Claim~\ref{real aug}, the real rank of the augmented matrix $A''$ is $R_{\mathbb{R}}(A'') = R_{\mathbb{R}}(A') = 3d$.
As to the boolean rank of $A''$, using Claim~\ref{diagonal} we can conclude that  $R_{bool}(A'') = 4d$,
and we have thus proved:

\BT
For every $d$, there exists a matrix $M$
such that $R_{bool}(M) = 4d$ and $R_{\mathbb{R}}(M) = 3d$.
\ET

We note that the same technique as used in the proof of Theorems 3 and 4,
yields a similar gap between the non-negative rank of $A''$ and the real rank of $A''$.
This follows easily by showing that the non-negative rank of the matrix $(A|x,y)$ is $4$, where $A$ is the matrix
in Equation~\ref{matrix for boolean gap} and $x = (1,0,0,1)^T, \ y = (1,1,0,0)^T$ are as above.
Indeed we have that $3 = R_{\mathbb{R}}(A|x,y) \leq R_{\mathbb{R}^+}(A|x,y) \leq R_{binary}(A|x,y) = 4$.
But Watson showed in~\cite{Watson}
that if the non-negative rank of a given binary matrix is bounded above by $3$,
then its binary rank is equal to its non-negative rank.
Hence if we assume that $R_{\mathbb{R}^+}(A|x,y) = 3$ then we get also that $R_{binary}(A|x,y) = 3$.
We can now continue as above and get again that for every $d$, there exists a binary matrix $M$
such that $R_{\mathbb{R}^+}(M) = 4d$ and $R_{\mathbb{R}}(M) = 3d$.

This example also shows that the non-negative rank does not always have the Augmentation property.

\section{Increasing the Binary and Boolean Rank}
\label{section increase rank}
In Section~\ref{gaps}, we constructed a matrix $A'$ of size
$ \Theta(d)\times \Theta(d)$ and binary/boolean rank $3d$, that had a constant size matrix $A$ on its diagonal.
Then by augmenting $A'$ with vectors that belong to sources of its base graph,
we were able to increase its binary/boolean rank to $4d$,
although when augmenting the matrix with each one of these vectors separately
the binary/boolean rank did not increase.

In this section we show that in fact we can get an unbounded gap between the rank of the augmented matrix
and the rank of the original matrix, for both the binary and boolean rank.

\subsection{Increasing the Binary Rank by Duplicating Rows}
\label{duplicate rows subsection}
We first consider the binary rank, and using
the understanding we gained about identical rows in matrices that do not posses
the Unique base rows sums property, we show how to
design a family of matrices in which there is an unbounded gap
between $R_{binary}(A|x_1,...,x_t)$ and $R_{binary}(A)$, although $R_{binary}(A|x_i) = R_{binary}(A)$
for $1 \leq i \leq t$.
Unfortunately, the real rank of the matrix $A$ used in this second construction,
also increases when augmenting $A$,
and thus in this case we do not get a gap between the real and binary rank.

Consider the following matrix that is similar to the one in Example~\ref{identical rows example}:

$$
\left(
     \begin{array}{cccc}
 1& 1&  1 & 1  \\
 1& 1&  1 & 1  \\
0& 0 & 1 &  1 \\
1& 1& 0& 0 \\
 1& 0 & 0 &  1 \\
 0&  1 & 1 &  0 \\

\end{array}
\right)
$$

The binary rank of this matrix is $4$, and it does not have the Unique base rows sums property.
Also, as in Example~\ref{identical rows example}, this matrix does not have the Augmentation property,
and thus if we augment it with vectors from bases that are sources in its base graph,
we can increase the binary rank by at least $1$.

Now consider the following matrix $A_k$, whose rank is also $4$,
but is a $(k+4)\times 4$ matrix, in which the top $k$ rows are identical:

\begin{equation}
\label{Ak matrix}
A_k = \left(
     \begin{array}{cccc}
 1& 1&  1 & 1  \\
 \vdots & & & \vdots \\
 1& 1&  1 & 1  \\
0& 0 & 1 &  1 \\
1& 1& 0& 0 \\
 1& 0 & 0 &  1 \\
 0&  1 & 1 &  0 \\
\end{array}
\right)
\end{equation}

We will show that we can choose vectors $x_1,\ldots,x_k$ such that
$R_{binary}(A_k|x_1,...,x_k) = R_{binary}(A_k)+ k-1 = 3+k$,
although $R_{binary}(A_k|x_i) = R_{binary}(A_k) = 4$ for $1\leq i \leq k$.
Hence, we are able to increase the binary rank of the augmented matrix from $4$ to $3+k$ for any $k$.

Let $y$ and $z$ be two $k$-dimensional binary vectors satisfying $y+z=(1,1,\ldots,1)^T$,
and let $B_{y,z}$ be the set of the following four vectors:
$$B_{y,z} =
 \left(
     \begin{array}{c}
      y \\
      1\\
      0\\
      0\\
      0\\
     \end{array}
   \right),
 \left(
     \begin{array}{c}
       y \\
       0\\
       1\\
       0 \\
      0\\
     \end{array}
   \right),
    \left(
     \begin{array}{c}
       z \\
         0\\
         0\\
       1\\
      0\\
     \end{array}
   \right),
     \left(
     \begin{array}{c}
       z \\
       0 \\
      0\\
      0\\
     1\\
     \end{array}
   \right)
$$
Then $B_{y,z}$ is a base of $A_k$, and thus in particular $R_{binary}(A_k|x) = R_{binary}(A_k)$
for every $x \in B_{y,z}$. Therefore, we get that:
\BL
Let $x$ by a $(k+4)$-dimensional binary vector that has exactly one nonzero entry in its last $4$
coordinates. Then $R_{binary}(A_k|x) = R_{binary}(A_k)$.
\EL
The above lemma gives an abundance of vectors that we can augment the matrix $A_k$ with,
without increasing its rank when $A_k$ is augmented with each of these vectors separately.
In fact, the first $k$ coordinates of these vectors are not restricted. This abundance makes it easy
to increase the rank of the augmented matrix when augmenting the matrix with a subset of these vectors simultaneously.

For example, we can augment $A_k$ with
the $k$ vectors $x_i = (e_i,1,0,0,0)^T$ for $i = 1,...,k$.
The resulting $(k+4)\times (k+4)$ matrix is:
\begin{equation}
\label{duplicate rows matrix}
(A_k|x_1,...,x_k) = \left(
     \begin{array}{cccc|ccccc}
 1& 1&  1 & 1 & 1 & 0 & 0  & 0 & 0  \\
  1& 1&  1 & 1 & 0 & 1 & 0  & 0  & 0  \\
   1& 1&  1 & 1  & 0 & 0 & 1 & 0   & 0 \\
 \vdots & & & \vdots & \vdots&   & & & \vdots \\
 1& 1&  1 & 1  & 0 &0 & 0 & 0   & 1  \\\hline
   0& 0&  1 & 1  &1&  1 & 1   & 1 & 1  \\
1& 1 & 0 &  0  & 0 &0& 0 & 0  & 0\\
 1& 0 & 0 &  1  & 0 &0& 0 & 0   & 0 \\
 0&  1 & 1 &  0  & 0&0 & 0 & 0   & 0 \\
\end{array}
\right)
\end{equation}

Using this matrix we can now prove the following theorem:
\BT
\label{duplicate theorem}
For any $k$, there exists a matrix $A_k$ and vectors $x_1,...,x_k$,
such that  $R_{binary}(A_k|x_i) =  R_{binary}(A_k) = 4$ for $1 \leq x_i \leq k$, but
$R_{binary}(A_k|x_1,...,x_{k}) =  k+3$.
\ET
\BPF
Denote by $A'_k$ the augmented matrix $(A_k|x_1,...,x_{k})$ presented above
in Equation~\ref{duplicate rows matrix}, and
note that $A'_k$ has the following structure:
$$
A'_k =
\left(
 \begin{array}{cc}
B_k  &   D_k \\
  C  & 0  \\
\end{array}
\right)
$$
where $D_k$ is a $(k+1) \times k$ matrix whose first $k$ rows are equal to $I_k$,
and the last row of $D_k$ is the all $1$ vector.

For a block matrix of this type, it holds that $R_{\mathbb{R}}(A'_k) \geq R_{\mathbb{R}}(C)+  R_{\mathbb{R}}(D_k)$.
It is easy to verify that $R_{\mathbb{R}}(C) = 3$, and that $R_{\mathbb{R}}(D_k) = k$. Thus, $R_{\mathbb{R}}(A'_k) \geq k+3$.

But the real rank is always bounded above by the binary rank,
and thus we get that  $R_{binary}(A'_k) \geq R_{\mathbb{R}}(A'_k) \geq k+3$.
We also have that $R_{binary}(A'_k) \leq k+3$, because $R_{binary}(A_k|x_1) = R_{binary}(A_k) = 4$,
and so $R_{binary}(A_k|x_1,...,x_{k}) \leq 4 + k-1 = k+3$. Hence, $R_{binary}(A'_k)=k+3$.
\EPF\\

We note that as opposed to Claim~\ref{real aug},
here we have that $R_{binary}(A) = 4$ whereas $R_{\mathbb{R}}(A) = 3$,
and as we saw in the proof of Theorem~\ref{duplicate theorem}, the real rank of the augmented
matrix also increased and did not stay constant. This is of course expected, since
otherwise we would get an unbounded gap between the real and binary rank, and it is known
that $R_{binary}(A) \leq 2^{O(\sqrt{R_{\mathbb{R}}(A)})}$ (See~\cite{Shachar}).

\subsection{Increasing the Boolean Rank by Duplicating Rows}
\label{Boolean rank duplicate}
The above construction can be used to prove a similar result for the boolean rank as
that given in Theorem~\ref{duplicate theorem} for the binary rank.

Consider the same matrix $A_k$ as in Equation~\ref{Ak matrix}.
Then it is easy to verify that $R_{bool}(A_k) = 4$.
Now we again augment $A_k$ with the vectors $x_i = (e_i,1,0,0,0)^T$ for $i = 1,...,k$,
and the resulting  $(k+4)\times (k+4)$ matrix is again as in Equation~\ref{duplicate rows matrix}.

Note that the augmented matrix has the matrix $I_k$ as a sub-matrix and thus
$R_{bool}(A_k|x_1,...,x_k) \geq R_{bool}(I_k)$. But $R_{bool}(I_k)=k$, since no two $1$'s on the diagonal
of $I_k$ can be in the same monochromatic rectangle.
We have thus proved:

\BT
For any $k$, there exists a matrix $A_k$ and vectors $x_1,...,x_{k}$,
such that  $R_{bool}(A_k|x_i) =  R_{bool}(A_k) = 4$ for $1 \leq x_i \leq k$, but
$R_{bool}(A_k|x_1,...,x_{k}) \geq k$.
\ET

\section{Conclusion and Open Problems}
\label{section open problems}
We have introduced the Augmentation property for matrices and gave a sufficient and necessary condition
for it to hold with respect to the binary and boolean rank of binary matrices. Furthermore,
we gave a characterization of binary matrices that have this property under the binary rank,
and used the augmentation property to construct a family of binary matrices with a linear gap between
the binary/boolean and the real rank. As we showed, when the Augmentation property does not
hold, there exists a family of binary matrices for which there is an unbounded gap between their binary/boolean rank
before and after augmenting them with a subset of vectors, although when augmenting them with each vector
separately the rank does not increase.

Many open questions remain. In particular, note that in our proof of Theorem 1, we stated that
the augmentation property does not hold when the base graph has two sources.
In this case we increased the rank of the matrix by augmenting it with {\bf all} vectors in the bases
represented by these two sources.
However, all our examples in this paper, and also the constructions presented in Section~\ref{duplicate rows subsection}
and Section~\ref{Boolean rank duplicate}, in fact use only {\bf one} vector from each source base.

Thus it is interesting to prove if in general when we want to increase the rank of a matrix
that does not have the augmentation property, it is indeed necessary to augment the matrix
with all vectors in the two sources, or there always exists one vector in each one of the two sources
that can be used to augment the matrix and increase its rank.


Furthermore, the characterization given in Theorem~\ref{7} for the binary rank, is a
characterization of the structure of the bases of a matrix, which guarantees the Augmentation property.
Is there a characterization of the structure of the matrix itself that guarantees the Augmentation property?
Also, as we saw, this characterization does not hold for the boolean rank. Is there an interesting characterization
for the boolean rank that guarantees the Augmentation property?

Finally, it would be of course interesting to use the technique presented in Section~\ref{gaps}
to get a larger gap between the binary/boolean rank and the real rank. To this end note that
in our construction we only used two bases $U_1,U_2$ of the original matrix $A$
to create a gap between the real and binary rank of $A'=I_d \otimes A$,
although the matrix $A'$ had an exponential number of bases.

It is of course also interesting to examine the Augmentation property and the concepts
defined in this paper for other rank functions, such as the non-negative rank (as we saw the non-negative rank
does not always have the Augmentation property), and continue the study of the differences between the various rank functions
for other properties that hold trivially for the real rank.

\bibliographystyle{plain}
\bibliography{augmentjournalfinal}

\end{document}